\documentclass[pra, twocolumn, letterpaper,superscriptaddress]{revtex4}

\usepackage{commands}

\begin{document}

\title{Wave kinetic equation in a nonstationary and inhomogeneous medium \\ with a weak quadratic nonlinearity}

\begin{abstract}
We present a systematic derivation of the wave kinetic equation describing the dynamics of a statistically inhomogeneous  incoherent wave field in a medium with a weak quadratic nonlinearity. The medium can be nonstationary and inhomogeneous. Primarily based on the Weyl phase-space representation, our derivation assumes the standard geometrical-optics ordering and the quasinormal approximation for the statistical closure. The resulting wave kinetic equation simultaneously captures the effects of the medium inhomogeneity (both in time and space) and of the nonlinear wave scattering. This general formalism can serve as a stepping stone for future studies of weak wave turbulence interacting with mean fields in nonstationary and inhomogeneous media.
\end{abstract}

\author{D.~E. Ruiz}
\author{M. E. Glinsky}
\affiliation{Sandia National Laboratories, P.O. Box 5800, Albuquerque, New Mexico 87185, USA}
\author{I.~Y.~Dodin}
\affiliation{Department of Astrophysical Sciences, Princeton University, Princeton, New Jersey 08544, USA}
\affiliation{Princeton Plasma Physics Laboratory, Princeton, New Jersey 08543, USA}

\date{\today}

\maketitle

%%%%%%%%%%%%%%%%%%%%%%%%%%%%%%%%%%%%%%%%%%%%%%
%%%%%%%%%%%%%%%%%%%%%%%%%%%%%%%%%%%%%%%%%%%%%%
\section{Introduction}
\label{sec:intro}

Wave turbulence is an ubiquitous phenomenon that has been studied in many contexts, e.g., water waves \cite{Zakharov:1992dg,Janseen:2004vj}, magnetohydrodynamic and other plasma waves \cite{sagdeev1969nonlinear,tsytovich2012nonlinear}, optical radiation \cite{Picozzi:2014bh}, and the dynamics of the Bose--Einstein condensate \cite{Lvov:2003cg,Nazarenko:2006co}. Many studies use the weak turbulence theory (WWT), which assumes that both the medium and the turbulent field are statistically homogeneous. (A modern review of WTT is given in \Ref{Nazarenko:2011cr}.) Under this assumption, a wave kinetic equation (WKE) is obtained that governs the dynamics of the wave-action density in the wavevector space $\vec{k}$ while the coordinate space $\vec{x}$ is considered as largely irrelevant. But this model fails to adequately capture the coherent structures that may arise from the small-scale turbulence, for example, stochastic solitons \cite{Picozzi:2011dv} and zonal flows in magnetized plasmas and atmospheres of rotating planets \cite{Lin:1998je,Terry:2000gw,Krommes:2002hva,Diamond:2004un,Fujisawa:2009jc,Parker:2014tb}. Regarding the latter in particular, the shearing of drift waves by zonal flows can lead to spectacular effects that are missed by homogeneous WTT, such as the suppression of drift-wave turbulence by zonal flows \cite{Lin:1998je,Rogers:2000ca,Terry:2000gw,Dimits:2000ki,StOnge:2017dta}. In order to retain these effects within the WWT framework, a more general theory is needed that would allow for inhomogeneous turbulence and consider the dynamics in the ``phase space'' $(\vec{x}, \vec{k})$.

To the extent that the nonlinear wave scattering can be neglected, the WKE is relatively easy to derive, at least in the geometrical-optics (GO) limit, in which case the WKE is simply the Liouville equation for the wave-action density in the $(\vec{x},\vec{k})$ space \cite{Tracy:2014to,McDonald:1985ib,McDonald:1991kk}. (In the GO limit, the latter is understood as the ray phase space.) In contrast, WKEs that simultaneously capture the inhomogeneities in the medium and nonlinear wave scattering have only been approached \textit{ad hoc}. For example, \Ref{Lvov:2003cg} makes use of Gabor wavelets to project the fluctuating wave field to the ($\vec{x},\vec{k})$ space. In statistical theories based on correlation functions \cite{Krommes:2002hva,Carnevale:2006id}, another approach consists in writing the correlation function in terms of slow and fast variables and then Fourier-transforming the fast variables. These approaches involve cumbersome calculations that are very specific to the problem of interest and hence hard to transfer from one problem to another.

%Previous WWTs have been developed for different dynamical systems. In order to obtain dynamical equations on the ray phase space ($\vec{x},\vec{k})$ (here, $\vec{x}$ is the spatial coordinate, and $\vec{k}$ is the wavevector coordinate), a variety of methods have been proposed. For example, \Ref{Lvov:2003cg} makes use of Gabor wavelets to project the fluctuating wave field to the ($\vec{x},\vec{k})$ space. In statistical theories based on correlation functions \cite{Krommes:2002hva,Carnevale:2006id}, another approach consists in writing the correlation function in terms of slow and fast variables and then Fourier-transforming the fast variables. These methods lead to WKEs describing the wave action density on the ray phase-space. However, these \textit{ad hoc} approaches involve tedious calculations that are very specific to the problem of interest and hence hard to transfer from one problem to another.

The aim of this paper is to construct a WKE for a general wave system with a quadratic nonlinearity. We shall assume that the medium in which the waves propagate is weakly nonstationary and inhomogeneous; \ie the characteristic period (temporal or spatial) of the waves is small compared to the inhomogeneity scale of the underlying medium. Under this assumption, it is possible to combine the well-known techniques of GO \cite{Whitham:2011kb,Tracy:2014to} with those frequently used in statistical descriptions of wave turbulence \cite{Nazarenko:2011cr,Picozzi:2014bh,frisch1995turbulence,kraichnan2013closure}.

The procedure that we shall employ is based on the \textit{Weyl symbol calculus} \cite{Weyl:1931uw,McDonald:1988dp} and consists of the following general steps: (i) we consider the turbulent wave field $\smash{\fluctpsi}(t,\vec{x})$ as a spacetime coordinate representation of an invariant vector $\smash{\ketlong{\fluctpsi}}$ in a Hilbert space; (ii) we write the corresponding partial-differential equation (PDE) for $\smash{\ketlong{\fluctpsi}}$; (iii) we introduce the correlation (or ``density'') operator $\smash{\avg{ \ketlong{\fluctpsi}\bralong{\fluctpsi}}}$ and derive the corresponding von Neumann equation by using a quasinormal approximation for the statistical closure; and (iv) we project this operator equation onto the $(\vec{x},\vec{k})$ space using the \textit{Weyl transform}. The resulting PDE is correct to all orders in the GO parameter \eq{eq:wke_epsilon}. Then, we take the GO limit using the well-known ordering hierarchy \cite{McDonald:1985ib,McDonald:1991kk} and reduce the PDE to the WKE [see \Eq{eq:wke}], which will capture both the turbulence inhomogeneity and the nonlinear wave scattering. The advantage of this approach is that it not only systematizes difficult calculations, which can lead to corrections to previous WKE models (see, \eg \Ref{Ruiz:2016gv}) but can also be extended to higher-order theories.

Notably, the Weyl calculus has already been used for developing inhomogeneous WTTs in optical wave turbulence, with new insights being gained on, \eg the incoherent modulational instability and incoherent solitons \cite{Hall:2002jc,Lisak:2003cm,Picozzi:2007dc,Picozzi:2008cc,Picozzi:2011dv,Picozzi:2014bha}. But unlike in those works, where the starting point is the nonlinear Schrödinger equation, here we are interested in WTT for general scalar wave systems that may not be manifestly quantumlike.

%We also emphasize that we seek to retain not only the effect of inhomogeneity but also the effect of the nonlinear wave scattering, which has been neglected so far in the Weyl analysis of turbulent dynamics. 

%It is worth mentioning that, in optical wave turbulence, the Weyl calculus has been quite successful for the development of inhomogeneous WTTs \cite{Hall:2002jc,Lisak:2003cm,Picozzi:2007dc,Picozzi:2008cc,Picozzi:2011dv,Picozzi:2014bha} with new insights being gained on, \eg the incoherent modulational instability \cite{Hall:2002jc} and incoherent solitons \cite{Picozzi:2011dv}. Contrary to these works, where the starting point is usually the nonlinear Schrödinger equation, here we are interested in WTT for general wave systems.

This work is organized as follows. In \Sec{sec:physical_model}, we introduce our basic equations, separate the wave field into its mean and fluctuating components, and obtain their respective governing equations. In \Sec{sec:stats}, we choose a statistical closure and obtain a closed equation for the correlation operator for the wave fluctuations. In \Sec{sec:WKE}, we introduce the Weyl transform, and we deduce the wave kinetic equation describing the dynamics of the wave fluctuations. In \Sec{sec:example}, we apply our general theory to a simple system as an example. Final comments and remarks are given in \Sec{sec:conclusions}. In \App{app:weyl}, we include a brief introduction to the Weyl calculus, which we extensively use in this paper. In \App{app:auxiliary}, we present some auxiliary calculations.

%%%%%%%%%%%%%%%%%%%%%%%%%%%%%%%%%%%%%%%%%%%%%%
%%%%%%%%%%%%%%%%%%%%%%%%%%%%%%%%%%%%%%%%%%%%%%
\section{Physical model}
\label{sec:physical_model}

%%%%%%%%%%%%%%%%%%%%%%%%%%%%%%%%%%%%%%%%%%%%%%
\subsection{Basic equations}
\label{sec:basic}

Let us consider a scalar real wave $\psi(t,\vec{x})$ propagating in a medium that is allowed to be nonstationary and inhomogeneous and also contains a second-order nonlinearity. Specifically, we assume $\psi$ to be governed by an equation of the form \cite{foot:nonlinear_interaction,foot:real_psi} 
\begin{equation}
	(\oper{D}_{\rm lin} \psi)(x) =
		(\oper{\alpha} \psi)(x)  \, (\oper{\beta} \psi)(x)
		+ \mc{S}(x).
	\label{eq:basic_original}
\end{equation}
Here $x \doteq (t,\vec{x})$ denotes the spacetime coordinate. (The symbol ``$\doteq$" stands for definitions.) Also, $\oper{D}_{\rm lin}$ is a linear operator, which can be represented as
\begin{equation}
	(\oper{D}_{\rm lin} \psi)(x) \doteq \int \mathrm{d}^4 x' \, \mcu{D}_{\rm lin}(x,x') \psi(x')  .
	\label{eq:basic_D_lin}
\end{equation}
The two-point scalar kernel $\mcu{D}_{\rm lin}(x,x')$ represents the linear-response model of the medium. The term $(\oper{\alpha} \psi) \, (\oper{\beta} \psi)$ is the quadratic nonlinearity in \Eq{eq:basic_original}, where $\oper{\alpha}$ and $\oper{\beta}$ are linear operators defined similarly as in \Eq{eq:basic_D_lin}. Finally, the term $ \mc{S}(x)$ in \Eq{eq:basic_original} represents a (possibly stochastic) source term.

Let us rewrite \Eq{eq:basic_original} as follows:
\begin{equation}
	(\oper{D}_{\rm lin} \psi)(x) =
		\frac{1}{2} f_{\rm nl}[\psi , \psi] (x)
		+  \mc{S}(x),
	\label{eq:basic_eq_sym}
\end{equation}
where $f_{\rm nl}[\phi,\psi](x)$ represents the nonlinear interaction,
\begin{equation}
	f_{\rm nl}[\phi,\psi] \doteq  
					(\oper{\alpha} \phi)  \,  (\oper{\beta} \psi) + 
					(\oper{\beta}  \phi)  \,  (\oper{\alpha} \psi) .
	\label{eq:basic_f_nl}
\end{equation}
Equation \eq{eq:basic_f_nl} is written in this manner so that $f_{\rm nl}[\phi,\psi](x)$ is symmetric with respect to its arguments; \ie
\begin{equation}
	f_{\rm nl}[\phi,\psi](x) = f_{\rm nl}[\psi,\phi](x).
	\label{eq:basic_symmetry}
\end{equation}
This property will be useful for future calculations.

%%%%%%%%%%%%%%%%%%%%%%%%%%%%%%%%%%%%%%%%%%%%%%
\subsection{Dynamics of the mean and fluctuating components}
\label{sec:mean_fluctuating}

In the following, we shall be interested in fields $\psi$ that are separable into a large-amplitude, low-frequency, coherent component $\bar{\psi}$ and a small-amplitude, high-frequency, fluctuating component $\tilde{\psi}$. We shall express $\psi$ as follows:
\begin{equation}
	\psi(x) = \bar{\psi}(x) + \epsilon \, \fluctpsi(x) , 
		\label{eq:basic_division}
\end{equation}
where $\bar{\psi} \doteq \avgsmall{ \psi }$ and $\avgsmall{ \cdot }$ denotes a statistical average. By definition, the fluctuating component of the wave field has a zero statistical average so $\avgsmall{ \fluctpsi }  =0$. Fluctuations in $\fluctpsi$ may be due to random initial conditions or due to a stochastic forcing term, which will be further discussed below. Also, $\epsilon \ll 1$ is some small parameter characterizing the amplitude of the fluctuations.

Inserting \Eq{eq:basic_division} into \Eq{eq:basic_eq_sym} leads to the governing equations for the mean and fluctuating fields
\begin{subequations}	\label{eq:basic_mean_fluct}
	\begin{align}
		(\oper{D}_{\rm lin} \bar{\psi}) (x)  
					&	=		\frac{1}{2} \,
								\big\{  f_{\rm nl}[\bar{\psi}, \bar{\psi} ]
											+ \epsilon^2  \avg{   f_{\rm nl} [\fluctpsi ,\fluctpsi ] }  \big\}
								+  \bar{ \mc{S}} , 
		\label{eq:basic_mean}	\\
		(\oper{D} \fluctpsi ) (x)  
					&	=		\frac{\epsilon}{2} \,
								\big\{  f_{\rm nl} [\fluctpsi ,\fluctpsi ]  
													- \avg{   f_{\rm nl} [\fluctpsi ,\fluctpsi ] }  \big\} 
								+ \epsilon  \, \widetilde{ \mc{S} } ,
		\label{eq:basic_fluct}
	\end{align}
\end{subequations}
where we introduced a modified dispersion operator for the fluctuating wave dynamics
\begin{align}
	(\oper{D} \fluctpsi ) 
			&		\doteq		(\oper{D}_{\rm lin} \fluctpsi ) 		
									- f_{\rm nl} [\bar{\psi} ,\fluctpsi ]  \notag \\
			&		=				(\oper{D}_{\rm lin} \fluctpsi )    
									 -	(\oper{\alpha} \bar{\psi})  \,  (\oper{\beta} \fluctpsi)
									 -	(\oper{\beta} \bar{\psi} ) \, (\oper{\alpha} \fluctpsi) ,
	\label{eq:basic_D_nl}
\end{align}
which depends linearly on the mean field $\bar{\psi}$. Thus, $\oper{D}$ includes a nonlinear coupling between the mean and fluctuating fields. In \Eqs{eq:basic_mean_fluct}, we separated $\mc{S}$ into its mean and fluctuating components so that $\mc{S} = \bar{ \mc{S} }  + \epsilon^2 \widetilde{ \mc{S} }$ assuming that the fluctuating component of the forcing term scales as $\epsilon^2$.

%%%%%%%%%%%%%%%%%%%%%%%%%%%%%%%%%%%%%%%%%%%%%%
\subsection{Abstract vector representation}
\label{sec:Dirac_abstract}

To facilitate future calculations, let us write \Eq{eq:basic_fluct} for the wave fluctuations in the abstract Hilbert space $\vec{L^2}(\mathbb{R}^4)$ of wave states with inner product \cite{Dodin:2014hw,Littlejohn:1993bd}
\begin{equation}
	\inner{ \phi \mid \psi } = \int \mathrm{d}^4 x \, \phi^*(x) \, \psi(x),
\end{equation}
where the integral is taken over $\mathbb{R}^4$. In this representation, $\psi(x)$ is written as $\psi(x)=\inner{x \mid \psi}$, where $\ketlong{x}$ are the eigenstates of the coordinate operators such that $\inner{ x \mid \widehat{t} \mid x'} =	t \, \delta^4( x - x')$, and $\inner{ x \mid \widehat{\vec{x}} \mid x'} =	\vec{x} \, \delta^4( x - x')$. Since $\fluctpsi(x)$ is real, then $\inner{\fluctpsi \mid x} = \inner{ x \mid \fluctpsi}$. Also, the temporal momentum (frequency) operator $\widehat{\omega}$ is given by $\inner{ x \mid \widehat{\omega} \mid x' } = i \pd_t \delta^4(x -x')$, and the spatial momentum (wavevector) operator $\widehat{\vec{k}}$ is $\inner{ x \mid \widehat{\vec{k}} \mid x' } = - i \pd_\vec{x} \delta^4(x -x')$ \cite{foot:wavevector}. In this representation, \Eq{eq:basic_fluct} can be rewritten as
\begin{equation}
	\oper{D}  \ketlong{ \fluctpsi }  
						=		\frac{\epsilon}{2} \,
								\big\{  \ketlong{ f_{\rm nl} [\fluctpsi ,\fluctpsi ] } 
													- \avg{  	\ketlong{ f_{\rm nl} [\fluctpsi ,\fluctpsi ] } } \big\}  
								+ \epsilon \ketlong{\widetilde{ \mc{S} }} .
	\label{eq:abstract_fluct}
\end{equation}
Here $\oper{D}$ is the modified dispersion operator such that $\inner{x \mid \oper{D} \mid \psi}	=		(\oper{D} \fluctpsi ) (x)$, where $(\oper{D} \fluctpsi ) (x)$ is given by \Eq{eq:basic_D_nl}. The ket $\ketlong{ f_{\rm nl}[\phi,\psi]}$ is given by
\begin{equation}
	\ketlong{ f_{\rm nl}[\phi,\psi]} 
		\doteq 	\int 	\mathrm{d}^4 x \, 	
					\ketlong{x} \inner{\phi \mid \oper{K} (x) \mid \psi } ,
	\label{eq:abstract_f_nl}
\end{equation}
where $\oper{K} (x)$ is the operator describing the nonlinear wave--wave interactions; namely,
\begin{equation}
	\oper{K}(x) 	\doteq	
							\oper{\alpha}_{\rm T}\ketlong{ x} \bralong{x}\oper{\beta} 
							+ \oper{\beta}_{\rm T}\ketlong{ x} \bralong{x}\oper{\alpha}	.
	\label{eq:abstract_K}
\end{equation}
Here $\oper{A}_{\rm T}$ denotes the transpose of a given operator $\oper{A}$. [For more details on this calculation of \Eqs{eq:abstract_f_nl} and \eq{eq:abstract_K}, see \App{app:construction}.] It is to be noted that \Eqs{eq:basic_fluct} and \eq{eq:abstract_fluct} are equivalent. Indeed, \Eq{eq:basic_fluct} can be considered as the coordinate representation of \Eq{eq:abstract_fluct}.

%%%%%%%%%%%%%%%%%%%%%%%%%%%%%%%%%%%%%%%%%%%%%%
%%%%%%%%%%%%%%%%%%%%%%%%%%%%%%%%%%%%%%%%%%%%%%
\section{Statistical closure}
\label{sec:stats}

%%%%%%%%%%%%%%%%%%%%%%%%%%%%%%%%%%%%%%%%%%%%%%
\subsection{Statistical closure problem}
\label{sec:stats_problem}

The correlation operator for the fluctuating fields is
\begin{equation}
	\oper{W} \doteq \avg{ \ketlong{ \fluctpsi} \bralong{ \fluctpsi }},
	\label{eq:stats_Wigner}
\end{equation}
which can be viewed as the statistical average of the density operator for the fluctuations. As a side note, the coordinate representation of $\oper{W}$ is the correlation function
\begin{align}
	\mcu{C}(t,\vec{x},t',\vec{x}') 
				&	\doteq \avg{ \fluctpsi(t,\vec{x}) \fluctpsi(t',\vec{x}')} 	\notag \\
				&	= 	\avg{ \inner{ x \mid \fluctpsi} \inner{ \fluctpsi \mid x' }}  \notag \\
				&	= 	\bralong{ x} \avg{ \ketlong{\fluctpsi} \bralong{ \fluctpsi } } \ketlong{ x' }   \notag \\
				&	=  \inner{ x \mid \oper{W} \mid x' }  ,
\end{align}
which is commonly used in statistical theories based on the second-order cumulant expansion. (For more information, see \eg \Refs{frisch1995turbulence,kraichnan2013closure,Krommes:2002hva} and references therein.) 

Let us write \Eq{eq:basic_mean} in terms of $\oper{W}$. First, we introduce the trace operation, which can be expressed as $\mathrm{Tr}( \, \oper{A}\,) = \int \mathrm{d}^4 x \, \inner{x \mid \oper{A} \mid x}$ for an arbitrary operator $\oper{A}$. Then,
\begin{align}
	\avg{ f_{\rm nl} [\fluctpsi ,\fluctpsi ](x) } 
		&	=	\avg{ \inner{ \fluctpsi \mid \oper{K}(x) \mid \fluctpsi} } \notag \\		
		&	=	\mathrm{Tr}[ \, \oper{K}(x)  \avg{ \ketlong{ \fluctpsi}	\bralong{\fluctpsi} } \, ]		\notag\\
		&	=	\mathrm{Tr}[ \, \oper{K}(x)  \oper{W} \, ] \notag \\
		&	=	\inner{ x \mid ( \oper{\beta} \oper{W} \oper{\alpha}_{\rm T}+ \oper{\alpha} \oper{W} \oper{\beta}_{\rm T}) \mid x} \notag \\
		&	=	2 \inner{x \mid \oper{\alpha} \oper{W} \oper{\beta}_{\rm T}) \mid x},
\end{align}
where we used \Eqs{eq:abstract_f_nl} and \eq{eq:stats_Wigner}. Equation \eq{eq:basic_mean} becomes
\begin{align}
		(\oper{D}_{\rm lin} \bar{\psi}) (x) 
					&	=		\frac{1}{2} \,
								 f_{\rm nl}[\bar{\psi}, \bar{\psi} ](x) 
								 + \epsilon^2 \inner{x \mid \oper{\alpha} \oper{W} \oper{\beta}_{\rm T} \mid x}  
								+  \bar{ \mc{S} } .
		\label{eq:stats_mean}
\end{align}
Thus, one sees that the correlation operator $\oper{W}$ acts as a source term in \Eq{eq:stats_mean} for the mean field $\bar{\psi}$. 

Now, let us obtain the governing equation for $\oper{W}$. Multiplying \Eq{eq:abstract_fluct} by $\bralong{\fluctpsi}$ from the right and averaging leads to an equation for the correlation operator:
\begin{align}
	\oper{D}  \oper{W}
				=	& \, 	\frac{\epsilon}{2}	
							\avg{ \ketlong{ f_{\rm nl} [\fluctpsi ,\fluctpsi ] } \bralong{\fluctpsi}    } 
					 + \epsilon \, \avg{ \ketlong{\widetilde{ \mc{S} }} \bralong{\fluctpsi} }.
	\label{eq:stats_eq_original}
\end{align}
Subtracting from \Eq{eq:stats_eq_original} its Hermitian conjugate gives
\begin{align}
		[\oper{D}_{\rm H}, \oper{W} ]_- 
		+ i [\oper{D}_{\rm A}, \oper{W} ]_+ 
			 	=	& \,	i \epsilon 
						\big[ 	
								\avg{   \ketlong{ f_{\rm nl} [\fluctpsi ,\fluctpsi ] }   \bralong{\fluctpsi}   }
						\big]_{\rm A}				\notag \\
					&	+ 2i \epsilon \,
						\big[ \avg{ \ketlong{\widetilde{ \mc{S} }} \bralong{\fluctpsi}  } \big]_{\rm A},
		\label{eq:stats_main}
\end{align}
where we separated the modified dispersion operator $\oper{D}$ into its Hermitian and anti-Hermitian parts so that $\oper{D}  = \oper{D}_{\rm H} + i \oper{D}_{\rm A}$. Here the Hermitian and anti-Hermitian parts of an arbitrary operator $\oper{A}$ are defined as $\oper{A}_{\rm H} \doteq (\oper{A} + \oper{A}^\dag)/2$ and $\oper{A}_{\rm A} \doteq (\oper{A} - \oper{A}^\dag)/(2i)$. Note that, both $\oper{D}_{\rm H}$ and $\oper{D}_{\rm A}$ are Hermitian by definition. Also, $[ \cdot , \cdot ]_{\mp}$ respectively denote the commutators and anticommutators; \ie $[\oper{A}, \oper{B}]_- = \oper{A} \oper{B} - \oper{B} \oper{A}$ and $[\oper{A}, \oper{B}]_+ = \oper{A} \oper{B} + \oper{B} \oper{A}$.

Equations \eq{eq:stats_mean} and \eq{eq:stats_main} are not closed. The left hand-side of \Eq{eq:stats_main} is written in terms of $\oper{W}$, which is bilinear in the fluctuating wave field, but the right-hand side of \Eq{eq:stats_main} contains terms that are linear and cubic with respect to the fluctuating field. This is the fundamental \textit{statistical closure problem} \cite{frisch1995turbulence,kraichnan2013closure,Krommes:2002hva}. The next step is to obtain a statistical closure in order to express \Eq{eq:stats_main} in terms of the correlation operator $\oper{W}$ only.

%%%%%%%%%%%%%%%%%%%%%%%%%%%%%%%%%%%%%%%%%%%%%%
\subsection{Quasilinear approximation}
\label{sec:QL}

One possible closure is to neglect the cubic interaction term in \Eq{eq:stats_main}. Then, one linearizes \Eq{eq:abstract_fluct} so that
\begin{equation}
	\oper{D} \ketlong{ \fluctpsi } \simeq \epsilon \ketlong{\widetilde{ \mc{S} }}.
\end{equation}
After formally inverting $\oper{D}$ and substituting into \Eq{eq:stats_main}, we obtain the following equation for $\oper{W}$:
\begin{equation}
	[\oper{D}_{\rm H}, \oper{W} ]_- + i [\oper{D}_{\rm A}, \oper{W} ]_+ 
		=	2i \epsilon^2  \,
			\big[ \oper{S} (\oper{D}^{-1})^\dag \big]_{\rm A},
	\label{eq:stats_quasilinear}
\end{equation}
where $\oper{S}	\doteq \avg{ \ketlong{\widetilde{ \mc{S} }} \bralong{\widetilde{ \mc{S} }} }$ is the density operator associated to the source fluctuations.  This approximation, where fluctuations are described by a linear equation [\Eq{eq:stats_quasilinear}] while the only nonlinearity is retained in the equation for the mean field [\Eq{eq:stats_mean}], is usually referred as the \textit{quasilinear approximation}. It has been used to study a wide variety of physical systems, \eg the emergence of zonal flows from Rossby-wave turbulence \cite{Srinivasan:2012im}, wave--particle interactions in plasmas \cite{Kaufman:1972ja,Dewar:1973jq}, and turbulence and dynamo induced by the magnetorotational instability \cite{Squire:2015fk}. Although the quasilinear approximation provides a valid statistical closure, the resulting WKE will not include the effects of nonlinear wave--wave  scattering. Since it is our goal to compute such nonlinear interactions, the cubic nonlinearity in \Eq{eq:stats_main} needs to be retained.

%%%%%%%%%%%%%%%%%%%%%%%%%%%%%%%%%%%%%%%%%%%%%%
\subsection{A statistical closure beyond the quasilinear approximation}
\label{sec:closure}

Unlike the quasilinear model discussed in \Sec{sec:QL}, here we shall adopt a statistical closure that perturbatively incorporates nonlinear effects into the equation for the fluctuating field. Let us separate $\fluctpsi$ into two parts:
\begin{equation}
	\ketlong{\fluctpsi} = \ketlong{\fluctpsi_0} + \epsilon \ketlong{\fluctpsi_1}.
	\label{eq:closure_separation}
\end{equation}
Here $\ketlong{\fluctpsi_0} = \mc{O}(1)$ is chosen to satisfy the linear part of \Eq{eq:abstract_fluct}; \ie $\oper{D} \ketlong{\fluctpsi_0}  =0$. The fluctuations in $\fluctpsi_0$ are due to random initial conditions, whose statistics are considered to be uncorrelated to those of $\widetilde{ \mc{S} }$. Regarding $\fluctpsi_1$, when formally inverting $\oper{D}$ in \Eq{eq:abstract_fluct}, one finds that to lowest order in $\epsilon$, $\fluctpsi_1$ is given by
\begin{equation}
	 \ketlong{ \fluctpsi_1 }  
				=	\ketlong{ \widetilde{\phi} }  	+ \oper{D}^{-1} \ketlong{ \widetilde{ \mc{S} }} ,
	\label{eq:closure_fluct}
\end{equation}
where
\begin{equation}
		\ketlong{ \widetilde{\phi} }  
				\simeq	\frac{1}{2} \,
							\oper{D}^{-1} 
						\big\{   \ketlong{ f_{\rm nl} [\fluctpsi_0 ,\fluctpsi_0 ] } 
				 					-	\avg{  \ketlong{ f_{\rm nl} [\fluctpsi_0 ,\fluctpsi_0 ] } } \big\} .
	\label{eq:closure_phi}		
\end{equation}
After substituting \Eqs{eq:closure_separation}--\eq{eq:closure_phi} into the right-hand side of \Eq{eq:stats_main}, we obtain
\begin{widetext}
\begin{align}
	[\oper{D}_{\rm H}, \oper{W} ]_- 	+ i [\oper{D}_{\rm A}, \oper{W} ]_+ 
		=	&		\, i \epsilon  \big\{ 	\avg{
								\ketlong{ f_{\rm nl} [\fluctpsi_0 ,\fluctpsi_0 ] }  
								\bralong{\fluctpsi_0}	}	\big\}_{\rm A}	
					+ 2i \epsilon^2  \big\{ 	\avg{
								\ketlong{ f_{\rm nl} [\widetilde{\phi} ,\fluctpsi_0 ] }  
								\bralong{\fluctpsi_0}	}	\big\}_{\rm A}	
				+ i \epsilon^2		\big\{	\avg{   
								 \ketlong{ f_{\rm nl} [\fluctpsi_0 ,\fluctpsi_0 ] }  
								 \bralong{\widetilde{\phi}} } \big\}_{\rm A}	\notag \\
			&		+ 2 i \epsilon^2   \big[  \oper{S}  (\oper{D}^{-1})^\dag \big]_{\rm A} 
			+ \mc{O}(\epsilon^3).
		\label{eq:closure_main_aux}
\end{align}
\end{widetext}
Here we neglected $\mc{O}(\epsilon^3)$ terms, \eg $\epsilon^3 \ketlong{ f_{\rm nl} [\fluctpsi_0 ,\widetilde{\phi} ] }  \bralong{\widetilde{\phi}} $. Note that the averages of quantities involving both $\fluctpsi_0$ and $\widetilde{ \mc{S} }$, such as $\epsilon \avgsmall {\ketlong{\widetilde{ \mc{S} }}\bralong{\fluctpsi_0 } }$ and $\epsilon^2 \avgsmall{ \ketlong{ f_{\rm nl} [\fluctpsi_0 , \fluctpsi_0 ] }  \bralong{\widetilde{ \mc{S} }} }$, are zero since the statistics of $\fluctpsi_0$ and $\widetilde{ \mc{S} }$ are independent.  Finally, note that the factor of two appearing in the second term on the right-hand side of \Eq{eq:closure_main_aux} is due to the symmetry property \eq{eq:basic_symmetry}.

Now, let us calculate explicitly the statistical average of the nonlinear terms in \Eq{eq:closure_main_aux}. To do this, we shall use the quasinormal approximation which expresses higher-order $(n\geq 3) $ statistical moments of $\fluctpsi_0$ in terms of the lower-order moments. Specifically, one has
\begin{equation}
	\avg{ \fluctpsi_0(x_1)  \fluctpsi_0(x_2) \fluctpsi_0(x_3)  } =0,
	\label{eq:closure_quasi_cubic}
\end{equation}
\begin{align}
	\avg{ \fluctpsi_0(x_1) & \fluctpsi_0(x_2) \fluctpsi_0(x_3) \fluctpsi_0(x_4) } \notag\\
		=	   &	\,		\avg{ \fluctpsi_0(x_1) \fluctpsi_0(x_2) } \avg{ \fluctpsi_0(x_3) \fluctpsi_0(x_4) }  \notag \\
				&		+	\avg{ \fluctpsi_0(x_1) \fluctpsi_0(x_3) } \avg{ \fluctpsi_0(x_2) \fluctpsi_0(x_4) } 	\notag \\
				&		+	\avg{ \fluctpsi_0(x_1) \fluctpsi_0(x_4) } \avg{ \fluctpsi_0(x_2) \fluctpsi_0(x_3) } .
	\label{eq:closure_quasi_quartic}
\end{align}
Here $x_i=(t_i, \vec{x}_i)$ denotes a spacetime coordinate. 

Upon using \Eq{eq:closure_quasi_cubic}, one finds that the first term appearing in the right-hand side of \Eq{eq:closure_main_aux} vanishes; \ie $\avg{\ketlong{ f_{\rm nl} [\fluctpsi_0 ,\fluctpsi_0 ] }  \bralong{\fluctpsi_0}	}=0$. After substituting \Eqs{eq:abstract_f_nl} and \eq{eq:closure_phi}, one obtains the following for the second term on the right-hand side of \Eq{eq:closure_main_aux}:
\begin{widetext}
\begin{align}
	\big\langle \! \big\langle		\, \,  \ketlong{ f_{\rm nl} [\widetilde{\phi} ,\fluctpsi_0 ] } &
	 		\bralong{\fluctpsi_0} 	\, \, \big\rangle \! \big\rangle	
			=		\int \mathrm{d}^4 x \, 
					\ketlong{x}	\avg{     \inner{ \widetilde{\phi}  \mid \oper{K}(x) \mid \fluctpsi_0}  \bralong{\fluctpsi_0}	}	\notag \\
	&		=		\frac{1}{2}
							\int 	\mathrm{d}^4 x \,  \mathrm{d}^4 y \, 	
							\ketlong{x}
							\avg{     
								\big[  \inner{ \fluctpsi_0 \mid \oper{K}^\dag (y) \mid \fluctpsi_0 } 
					 					-	\avg{ \inner{ \fluctpsi_0 \mid \oper{K}^\dag(y) \mid \fluctpsi_0} } \big]		
								\inner{ y \mid (\oper{D}^{-1})^\dag \,	 \oper{K}(x) \mid \fluctpsi_0}  \bralong{\fluctpsi_0}	}	,
	\label{eq:closure_first}	
\end{align}
where $\ketlong{y}$ is an eigenstate of the coordinate operator. When using \Eq{eq:closure_quasi_quartic} in the abstract representation, one obtains
\begin{align}
	\avg{   \, \,  \big[  \inner{ \fluctpsi_0 \mid \oper{K}^\dag (y) \mid \fluctpsi_0} &
					 					-	\avg{ \inner{ \fluctpsi_0 \mid \oper{K}^\dag(y) \mid \fluctpsi_0} } \big]		
								\inner{ y \mid (\oper{D}^{-1})^\dag \,	 \oper{K}(x) \mid \fluctpsi_0}  \bralong{\fluctpsi_0}	\, \, }	
				\notag \\
		&	=	\contraction{\langle \, \,}{\fluctpsi_0}{\mid \oper{K}^\dag(y) \mid \fluctpsi_0 \rangle \langle  \rangle \, y \mid (\oper{D}^{-1})^\dag \,	 \oper{K}(x) \mid  }{ \fluctpsi_0}
				\contraction[2ex]{\langle \, \fluctpsi_0 \mid \oper{K}^\dag(y) \mid }{\fluctpsi_0 \,\,}{ \, \rangle \,\,\, \langle \,  \mid (\oper{D}^{-1})^\dag \,	 \oper{K}(x) \mid \fluctpsi_0 \, \rangle \, \, \, \langle \, }{ \fluctpsi_0}
				\inner{ \fluctpsi_0 \mid \oper{K}^\dag(y) \mid \fluctpsi_0}
				\inner{ y \mid (\oper{D}^{-1})^\dag \,	 \oper{K}(x) \mid \fluctpsi_0}  \bralong{\fluctpsi_0}	
				+
				\contraction{\langle \, \fluctpsi_0 \mid \oper{K}^\dag(y) \mid \,}{\fluctpsi_0 }{ \, \rangle \,\,\,\, \langle \,  \mid (\oper{D}^{-1})^\dag \,	 \oper{K}(x) \mid \,\, }{ \fluctpsi_0}
				\contraction[2ex]{\langle \, \,}{\fluctpsi_0}{\mid \oper{K}^\dag(y) \mid \fluctpsi_0 \rangle \langle \, \rangle \, y \mid (\oper{D}^{-1})^\dag \,	 \oper{K}(x) \mid  \fluctpsi_0 \, \rangle  \langle \, }{ \fluctpsi_0}				
				\inner{ \fluctpsi_0 \mid \oper{K}^\dag(y) \mid \fluctpsi_0}
				\inner{ y \mid (\oper{D}^{-1})^\dag \,	 \oper{K}(x) \mid \fluctpsi_0}  \bralong{\fluctpsi_0}	
				\notag \\
		&	= 2 
				\contraction{\langle \, y \mid (\oper{D}^{-1})^\dag \oper{K}(x) \mid \,\, }{\fluctpsi_0}{ \rangle \, \,  \langle \, }{ \fluctpsi_0}
				\contraction{\inner{ y \mid (\oper{D}^{-1})^\dag \,	 \oper{K}(x) \mid \fluctpsi_0} \langle \, \fluctpsi_0 \mid  \oper{K}(y) \mid \,\,}{\fluctpsi_0 }{ \, \langle \, \,   \rangle \,  }{ \fluctpsi_0}
				\inner{ y \mid (\oper{D}^{-1})^\dag \,	 \oper{K}(x) \mid \fluctpsi_0} 
				\inner{ \fluctpsi_0 \mid \oper{K}^\dag(y) \mid \fluctpsi_0}
				\bralong{\fluctpsi_0}
				\notag \\
		&	=	2	\bralong{y} \,	(\oper{D}^{-1})^\dag \, \oper{K}(x) \,
					\avg{ \ketlong{\fluctpsi_0}  \bralong{\fluctpsi_0} }
					 \, \oper{K}^\dag(y) \, 
					 \avg{ \ketlong{\fluctpsi_0}  \bralong{\fluctpsi_0} }
					\notag \\
		&	=	2	\bralong{y} 	\, (\oper{D}^{-1})^\dag\,  \oper{K}(x)  \, \oper{W}_0  \, \oper{K}^\dag(y) \, \oper{W}_0 ,
\end{align}
where the overlines denote the statistical averaging among the various wave field states. In the third line, we used the symmetry property \eq{eq:basic_symmetry}. Also, $\oper{W}_0 \doteq \avg{ \ketlong{\fluctpsi_0}  \bralong{\fluctpsi_0} }$. Substituting this result into \Eq{eq:closure_first} gives
\begin{equation}
	\avg{ \ketlong{ f_{\rm nl} [\widetilde{\phi} ,\fluctpsi_0 ] }  \bralong{\fluctpsi_0} }
		=		 \int  \mathrm{d}^4 x \,	\mathrm{d}^4 y \,
							\ketlong{x} \bralong{y}
							(\oper{D}^{-1})^\dag\,
							\oper{K}(x) \,
							\oper{W}_0 \,
							\oper{K}^\dag (y) \,
							\oper{W}_0	.		
	\label{eq:closure_res_I}	
\end{equation}
In a similar manner, substituting \Eq{eq:closure_phi} into the third term in \Eq{eq:closure_main_aux} leads to
\begin{align}
	\avg{   & \ketlong{ f_{\rm nl}  [\fluctpsi_0 ,\fluctpsi_0 ] } 	 \bralong{\widetilde{\phi}} }
			=	\int \mathrm{d}^4 x \, \ketlong{x} \avg{ 
					\inner{ \fluctpsi_0 \mid \oper{K}(x) \mid \fluctpsi_0}
					\bralong{\widetilde{\phi}} }
				\notag \\
		&	\quad \quad =	\frac{1 }{2}	 \int 	\mathrm{d}^4 x \, \mathrm{d}^4 y \, 	
							\ketlong{x}
							\bralong{y}  (\oper{D}^{-1})^\dag 
							\avg{	\inner{ \fluctpsi_0 \mid \oper{K}(x) \mid \fluctpsi_0}
							\big[  \inner{ \fluctpsi_0 \mid \oper{K}^\dag (y) \mid \fluctpsi_0} 
					 					-	\avg{ \inner{ \fluctpsi_0 \mid \oper{K}^\dag(y) \mid
					 									 \fluctpsi_0} } \, \big] \, }.
	\label{eq:closure_second}				
\end{align}
As before, when using the quasinormal approximation, we obtain
\begin{align}
	\big\langle \! \big\langle		\, \, \inner{   \fluctpsi_0 \mid \oper{K}(x) \mid \fluctpsi_0 } &
			\, \big[ \, \inner{ \fluctpsi_0 \mid \oper{K}^\dag (y) \mid \fluctpsi_0} 
			-	\avg{ \inner{ \fluctpsi_0 \mid \oper{K}^\dag(y) \mid \fluctpsi_0} } \, \big] \, \,  \big\rangle \! \big\rangle
				\notag \\
		&	=	\contraction{\langle \,\,}{\fluctpsi_0}{\mid \oper{K}(x) \mid \fluctpsi_0 \rangle \langle \, \,\, }{ \fluctpsi_0}
				\contraction[2ex]{\langle \, \fluctpsi_0 \mid \oper{K}(x) \mid \, }{\fluctpsi_0 }{ \, \rangle \,\,\,\, \langle \, \fluctpsi_0 \mid \oper{K}(y) \mid }{ \fluctpsi_0}
				\inner{ \fluctpsi_0 \mid \oper{K}(x) \mid \fluctpsi_0}
				\inner{ \fluctpsi_0 \mid \oper{K}^\dag (y) \mid \fluctpsi_0} 
		+
				\contraction{\langle \, \fluctpsi_0 \mid \oper{K}(x) \mid }{\fluctpsi_0 \, \, }{ \rangle \langle \, \,}{ \fluctpsi_0}
				\contraction[2ex]{\langle \, \,}{\fluctpsi_0 }{ \mid \oper{K}(x) \mid \fluctpsi_0 \, \rangle   \,\,\,\, \langle \, \fluctpsi_0 \mid \oper{K}(y) \mid }{ \fluctpsi_0}
				\inner{ \fluctpsi_0 \mid \oper{K}(x) \mid \fluctpsi_0}
				\inner{ \fluctpsi_0 \mid \oper{K}^\dag (y) \mid \fluctpsi_0} 
				\notag \\
		&	= 2 \contraction{\langle \, \fluctpsi_0 \mid \oper{K}(x) \mid }{\fluctpsi_0 \, \, }{ \rangle \langle \, \,}{ \fluctpsi_0}
				\contraction[2ex]{\langle \, \,}{\fluctpsi_0 }{ \mid \oper{K}(x) \mid \fluctpsi_0 \, \rangle   \,\,\,\, \langle \, \fluctpsi_0 \mid \oper{K}(y) \mid }{ \fluctpsi_0}
				\inner{ \fluctpsi_0 \mid \oper{K}(x) \mid \fluctpsi_0}
				\inner{ \fluctpsi_0 \mid \oper{K}^\dag (y) \mid \fluctpsi_0}
				\notag \\
		&	=	2	\, \mathrm{Tr} [ \, \oper{K}(x)  \, \avg{ \ketlong{\fluctpsi_0} \bralong{\fluctpsi_0} } \, \oper{K}^\dag(y)  \,
											\avg{ \ketlong{\fluctpsi_0}  \bralong{\fluctpsi_0} }	\,]
				\notag \\
		&	= 	2 \,	\mathrm{Tr} [ \, \oper{K}(x) \, \oper{W}_0 \, \oper{K}^\dag(y) \, \oper{W}_0 \, ] .
\end{align}
It is to be noted that the trace $\mathrm{Tr} [ \, \oper{K}(x) \, \oper{W}_0 \, \oper{K}^\dag(y) \, \oper{W}_0 \, ]$ is not an operator but rather a function of $x$ and $y$. Upon inserting this result into \Eq{eq:closure_second}, we obtain

\begin{equation}
	\avg{    \ketlong{ f_{\rm nl} [\fluctpsi_0 ,\fluctpsi_0 ] }  \bralong{\widetilde{\phi}} }
		=  \int 	\mathrm{d}^4 x \, \mathrm{d}^4 y \,
				 \ketlong{x} \bralong{y} (\oper{D}^{-1})^\dag  \,
				 \mathrm{Tr} [ \, \oper{K}(x) \, \oper{W}_0 \, \oper{K}^\dag(y) \, \oper{W}_0 \, ] .
		\label{eq:closure_res_II}		
\end{equation}
\end{widetext}
We then substitute \Eqs{eq:closure_res_I} and \eq{eq:closure_res_II} into \Eq{eq:closure_main_aux}. Replacing $\oper{W}_0$ with $\oper{W}$, which is valid to the leading order of accuracy of the theory, gives a closed equation for the correlation operator:
\begin{align}
		[\oper{D}_{\rm H}, \oper{W} ]_- 	+ i  [\oper{D}_{\rm A}, \oper{W} ]_+ 
			=	&	\, 2 i \epsilon^2	\big[ \oper{F} \,  (\oper{D}^{-1})^\dag \big]_{\rm A}  
					- 2 i \epsilon^2 	\big[ \widehat{\eta} \, \oper{W} \big]_{\rm A} 
					\notag \\				
				&	+ 2 i \epsilon^2 	\big[  \oper{S} (\oper{D}^{-1})^\dag \big]_{\rm A} ,
		\label{eq:closure_main}
\end{align}
where
\begin{subequations}	\label{eq:closure_scattering}
\begin{align}
	\widehat{\eta}	&\doteq	-
									\int 	\mathrm{d}^4 x \, 	\mathrm{d}^4 y \, 	
									\ketlong {x}	\bralong{y}
									(\oper{D}^{-1})^\dag
									\oper{K}(x) \, \oper{W} \, \oper{K}^\dag(y)   	,
				\label{eq:closure_eta}			\\			
	 \oper{F}		&\doteq
	 								\frac{1}{2}
	 								\int 	\mathrm{d}^4 x \, 	\mathrm{d}^4 y \, 	
									\ketlong {x}\bralong{y} 
									\mathrm{Tr} [ \, \oper{K}(x) \, \oper{W} \, \oper{K}^\dag (y) \, \oper{W} \, ]  		.
				\label{eq:closure_F}		 
\end{align}
\end{subequations}
For future purposes, note that $\oper{F}$ is Hermitian.

In summary, \Eqs{eq:stats_mean}, \eq{eq:closure_main}, and \eq{eq:closure_scattering} form a complete set of equations for the mean field $\bar{\psi}(x)$ and for the correlation operator $\oper{W}$ of the fluctuating wave field. In the following, we shall project \Eq{eq:closure_main} into the phase space by using the Weyl transform. Then, by adopting approximations that are consistent with the GO description of eikonal fields, we shall obtain a WKE describing the wave fluctuations. (Readers who are not familiar with the Weyl calculus are encouraged to read \App{app:weyl} before continuing further.)

%%%%%%%%%%%%%%%%%%%%%%%%%%%%%%%%%%%%%%%%%%%%%%
%%%%%%%%%%%%%%%%%%%%%%%%%%%%%%%%%%%%%%%%%%%%%%
\section{Dynamics in the ray phase space}
\label{sec:WKE}

%%%%%%%%%%%%%%%%%%%%%%%%%%%%%%%%%%%%%%%%%%%%%%
\subsection{Wigner--Moyal equation}

The Weyl transform is a mapping from operators in a Hilbert space into functions of phase space. In this work, the Weyl transform is defined as
\begin{equation}
	A(x,k) 	\doteq	\msf{W}[ \, \oper{A} \, ] = \int 	\mathrm{d}^4 s \,
									 	e^{i k \cdot s} 
									 	\inner{ x + \tfrac{1}{2} s  \mid \oper{A} \mid x - \tfrac{1}{2} s },
	\label{eq:wke_weyl}
\end{equation}
where $k\doteq (\omega, \vec{k})$ is the four wavevector, $s\doteq(\tau,\vec{s}) $, $k \cdot s = \omega \tau -  \vec{k} \cdot \vec{s}$, and $\mathrm{d}^4 s \doteq \mathrm{d} \tau \, \mathrm{d}^3 \vec{s}$. Here the integrals span over $\mathbb{R}^4$. We note that the \textit{Weyl symbol} $A(x,k) $ of an operator $\oper{A}$ is a function on the extended eight-dimensional phase space $z \doteq (x,k) = (t,\vec{x},\omega, \vec{k})$; \ie $A=A(t,\vec{x},\omega, \vec{k})$. Physically, the Weyl symbol $A(z)$ can be interpreted as a local Fourier transform of the spacetime representation of the operator $\oper{A}$ [see, for instance, \Eq{eq:weyl_weyl_x_rep}]. 

We shall refer to the Weyl symbol $W(z)$ corresponding to $\oper{W}$ as the \textit{Wigner function} of the fluctuating wave field \cite{foot:Wigner}. Since $\oper{W}$ is Hermitian, $W(z)$ is real. Moreover, since $\fluctpsi$ is real, one can write $W(z)$ as
\begin{equation}
	W(x,k) 			\doteq	\int 	\mathrm{d}^4 s \,
									 		e^{i k \cdot s} 
											\avg{	 	\fluctpsi( x + \tfrac{1}{2} s ) \,
												\fluctpsi( x - \tfrac{1}{2} s  ) }, 
\end{equation}
which implies that $W(x,k)=W(x,-k)$. Similar arguments apply to the Weyl symbol $S(z)$ corresponding to the operator $\oper{S} = \avg{ \ketlong{\widetilde{ \mc{S} }} \bralong{\widetilde{ \mc{S} }} }$.

The Weyl transform to \Eq{eq:closure_main} is given by
\begin{align}
		\moysin{ D_{\rm H}, W }	+	&    \moycos{D_{\rm A}, W} \notag \\
			=	& \, 2 \epsilon^2 \, \mathrm{Im} \left\{ F \star  [D^{-1}]^*  \right\}	 
				- 2 \epsilon^2 \, \mathrm{Im} \left( \eta \star  W \right)		 	\notag \\
				& + 2 \epsilon^2 \, \mathrm{Im}  \left\{ S \star  [D^{-1}]^* \right\} .
		\label{eq:wke_wigner_moyal}
	\end{align}
Here $D(z)$ is the Weyl symbol corresponding to $\oper{D}$. [In many cases, $D(z)$ is fairly straightforward to calculate.] From the properties of the Weyl transform, we observe that $D_{\rm H}(z) = \mathrm{Re} \, D$ and $D_{\rm A}(z)= \mathrm{Im} \, D$ are the (real) Weyl symbols corresponding to the operators $\oper{D}_{\rm H}$ and $\oper{D}_{\rm A}$, respectively. (``$\mathrm{Re}$" and ``$\mathrm{Im}$" denote the real and imaginary parts, respectively.) Likewise, $F(z)$, $\eta(z)$, and $[D^{-1}]^*(z)$ are the Weyl symbols corresponding to $\oper{F}$, $\oper{\eta}$, and $(\oper{D}^{-1})^\dag$, respectively. (These will be calculated explicitly later.) Also, ``$\star$'' is the Moyal product \eq{eq:weyl_Moyal}, and the brackets $\moysin{ \cdot , \cdot }$ and $\moycos{\cdot , \cdot }$ are the Moyal brackets [\Eqs{eq:weyl_sine_bracket} and \eq{eq:weyl_cosine_bracket}]. Basic properties of the Moyal product and the Moyal brackets are given in \App{app:weyl}.

Modulo the statistical closure introduced in \Sec{sec:stats}, \Eq{eq:wke_wigner_moyal} is an exact equation for the dynamics of the Wigner function $W(z)$. However, \Eq{eq:wke_wigner_moyal} is generally a pseudo-differential equation; \ie it involves an infinite number of partial derivatives, which appear in the Moyal products. This often makes \Eq{eq:wke_wigner_moyal} difficult to solve, both analytically and numerically. Because of this, we present below a simplification of this equation based on the GO approximation.

%%%%%%%%%%%%%%%%%%%%%%%%%%%%%%%%%%%%%%%%%%%%%%
\subsection{Wave kinetic equation}
\label{sec:wke_derivation}

Let us assume that the fluctuating fields have a characteristic temporal and spatial periods smaller than the characteristic time and length scales of the medium (described by $D_{\rm H}$, $D_{\rm A}$, and $S$). In other words, we consider
\begin{equation}
	\epsilon_{\rm go} 	
					\doteq 
					\mathrm{max} \left\{    \frac{1}{p_0 T}, \frac{1}{|\vec{p}| L} \right\} \ll 1,
	\label{eq:wke_epsilon}
\end{equation}
where $p_0$ and $\vec{p}$ are the characteristic frequency and wavenumber of the fluctuations and $L$ and $T$ are the characteristic time scale and length scale of the medium. The small parameter $\epsilon_{\rm go}$ is identified as the GO parameter. We shall also assume that $D(z)$ is only slightly dissipative so that the ordering $D_{\rm A} \to \epsilon_{\rm go} D_{\rm A} $ applies. These standard assumptions are frequently used in wave theories involving the GO approximation \cite{Whitham:2011kb,Tracy:2014to}.

To derive the WKE, let us first find a proper ansatz for $W(z)$. Note that \Eq{eq:stats_eq_original} reads as $\oper{D}_{\rm H} \oper{W} \simeq 0$ to lowest order. In the Weyl representation, this equation is written as $D_{\rm H} \star W \simeq 0$. Approximating the Moyal product gives to the lowest order
\begin{equation}
	D_{\rm H}(z) W(z) \simeq 0 .
	\label{eq:wke_dispersion}
\end{equation}
In order to satisfy \Eq{eq:wke_dispersion}, either $D_{\rm H}(z)$ or  $W(z)$ must be zero. Following \Refs{McDonald:1985ib,McDonald:1991kk}, we adopt the GO ansatz
\begin{equation}
	W(t,\vec{x},\omega,\vec{k}) =  2 \pi \delta \boldsymbol{(} D_{\rm H}(t,\vec{x},\omega,\vec{k}) \boldsymbol{)} J (t,\vec{x},\vec{k})	 ,
	\label{eq:wke_ansatz}
\end{equation}
where $J(t,\vec{x},\vec{k})$ is the on-shell distribution function or the wave-action density in the six-dimensional phase space.

As discussed in \Ref{McDonald:1985ib}, \Eq{eq:wke_ansatz} means that quanta of the wave fluctuations are localized on the \textit{dispersion manifold} defined by $D_{\rm H}(z)=0$ \cite{foot:reality}. This relation can be, in principle, inverted so that the wave frequency can be expressed as $\omega = \Omega(t,\vec{x},\vec{k})$. In general, there can be multiple solutions to $D_{\rm H}(z)$=0. Then, a different wave-action density function $J_\alpha(t,\vec{x},\vec{k})$ could be assigned to each mode $\alpha$. (If the dispersion curves approach each other closely enough such that mode conversion becomes possible, the GO ansatz \eq{eq:wke_ansatz} becomes inapplicable. Then, a more complicated approach is required \cite{Friedland:1987um,Flynn:1994cn,Dodin:2017kh,Ruiz:2017ij,Ruiz:2017wc}. This case is not consider here.)

%Regarding the assumed smoothness of the wave-action density $J (t,\vec{x},\vec{k})$, \Ref{McDonald:1985ib} argues that a large number of incoherent waves densely packed around a particular phase-space region could eventually lead to a slowly varying $J (t,\vec{x},\vec{k})$. Alternatively, \Ref{Ruiz:2017wc} proposes that a phase-space average or coarse graining of a Wigner function composed by many incoherent waves could eliminate fluctuations and lead to a smooth $J (t,\vec{x},\vec{k})$. In light of these arguments, it seems that $J (t,\vec{x},\vec{k})$ represents the distribution in phase-space of a wave bath of incoherent waves.

Let us now assume that the phase-space derivatives acting on $J$, $D_{\rm H}$, $D_{\rm A}$, $F$, $\eta$, and $S$ in \Eq{eq:wke_wigner_moyal} scale as $\pd_t \sim T^{-1}$, $\pd_\vec{x}\sim L^{-1}$, $\pd_{\omega} \sim p_0^{-1}$, and $\pd_\vec{k} \sim |\vec{p}|^{-1}$. For the sake of simplicity, we shall also consider that the medium inhomogeneities and the wave nonlinearities scale as follows: $\epsilon_{\rm go} \sim \epsilon^2$. With the previous assumptions, we substitute \Eq{eq:wke_ansatz} into \Eq{eq:wke_wigner_moyal}, and we integrate over the wave frequency $\omega$. Following the auxiliary calculations in \App{app:simplification}, one can approximate the Moyal products using integration by parts. Then, to lowest order in $\epsilon_{\rm go}$, \Eq{eq:wke_wigner_moyal} is given by
\begin{widetext}
\begin{equation}
	\int \mathrm{d} \omega \, 
		\big\{ D_{\rm H} \overleftrightarrow{\mc{L}} [  \delta(D_{\rm H}) J ]  	+ 2 D_{\rm A} J \delta(D_{\rm H} )  \big\} 
		=	\int \mathrm{d} \omega \,	\delta(D_{\rm H} )
				\left[ F - 2 \, \mathrm{Im} \left( \eta  \right)  J +   S \right]  + \mc{O}(\epsilon^2 ).
	\label{eq:wke_aux}
\end{equation}
\end{widetext}
where $\overleftrightarrow{\mc{L}}$ is the Janus operator \eq{eq:weyl_poisson_braket}. Upon using $[D^{-1}]  = D^{-1} + \mc{O}(\epsilon_{\rm go})$ \cite{foot:inverse} and the Sokhotski--Plemelj theorem \cite{foot:Plemelj}, we replaced $[D^{-1}]^* $ in \Eq{eq:wke_wigner_moyal} with its limiting form as $D_{\rm A}$ tends to zero:
\begin{equation}
	[D^{-1}]^* 
				\simeq \frac{1}{D_{\rm H} - i D_{\rm A} } 
				\to 	i \pi \delta(D_{\rm H}) 
						+ \mc{P} \, \frac{1}{D_{\rm H}}  ,
	\label{eq:wke_D}
\end{equation}
where ``$\mc{P}$" denotes the Cauchy principal value. Also, note that $F(z)$ and $S(z)$ are real because the corresponding operators are Hermitian. 

Upon integrating over the frequency, substituting \Eq{simp:eq:advection}, and neglecting higer-order terms in $\epsilon$, we obtain the nonlinear \textit{wave kinetic equation}
\begin{equation}
	\pd_t J + \{ J , \Omega \}	 
		=	2 \gamma J 
			+ 	S_{\rm ext} 
			+	C[J,J],
	\label{eq:wke}
\end{equation}
where $\{ \cdot , \cdot \} = \overleftarrow{\pd_\vec{x}}\cdot \overrightarrow{\pd_\vec{k}} - \overleftarrow{\pd_\vec{k}} \cdot \overrightarrow{\pd_\vec{x}}$ is the canonical six-dimensional Poisson bracket and 
\begin{subequations}
	\begin{align}
		\gamma(t, \vec{x},\vec{k}) 
			&	\doteq	-\left(	\frac{  D_{\rm A} }{ \pd D_{\rm H} / \pd \omega	} \right)_{\omega=\Omega} (t,\vec{x},\vec{k}) ,
			 \label{eq:wke_gamma} \\
		S_{\rm ext}(t, \vec{x},\vec{k}) 
			&	\doteq	\left(	\frac{  S }{ \pd D_{\rm H} / \pd \omega	} \right)_{\omega=\Omega} (t,\vec{x},\vec{k}) .
			 \label{eq:wke_source} 
	\end{align}
\end{subequations}
The nonlinear term $C[J,J](t,\vec{x},\vec{k})$ in \Eq{eq:wke} is
\begin{equation}
	C[J,J](t, \vec{x},\vec{k})	\doteq 	S_{\rm nl}[J,J] - 2\gamma_{\rm nl}[J]  J ,
	\label{eq:wke_scattering}
\end{equation}
where $\gamma_{\rm nl}[J] (t, \vec{x},\vec{k}) $ and $S_{\rm nl}[J,J](t, \vec{x},\vec{k})$ are given by
\begin{subequations}
	\begin{align}
		\gamma_{\rm nl}[J] (t, \vec{x},\vec{k}) 
			&	\doteq	\left(	\frac{ \mathrm{Im} \, \eta  }{ \pd D_{\rm H} / \pd \omega	} \right)_{\omega=\Omega} (t,\vec{x},\vec{k}),
			 \label{eq:wke_gamma_nl_aux} \\
		S_{\rm nl}[J,J](t, \vec{x},\vec{k}) 
			&	\doteq	\left(	\frac{  F  }{ \pd D_{\rm H} / \pd \omega	} \right)_{\omega=\Omega} (t,\vec{x},\vec{k}) . 
			 \label{eq:wke_source_nl_aux} 	
	\end{align}
\end{subequations}
As shown in \App{app:nonlinear}, to the leading order in $\epsilon_{\rm go}$, $\gamma_{\rm nl}[J] (t, \vec{x},\vec{k}) $ and $S_{\rm nl}[J,J](t, \vec{x},\vec{k}) $ are  given by
\begin{widetext}
	\begin{subequations}	\label{eq:wke_coeff_nl}
		\begin{align}
			\gamma_{\rm nl}[J] (t, \vec{x},\vec{k}) 
						&  =	-		\int \frac{\mathrm{d}^3 \vec{p} \, \mathrm{d}^3 \vec{q} }{(2 \pi)^3} \, 
									\delta^3(\vec{k} -\vec{p} -\vec{q}) \,
									\frac{\Theta(t,\vec{x},\vec{k},\vec{p},\vec{q})}{	\mc{N} } \,	
									\mathrm{Re} [ \, M(t,\vec{x}, \vec{p}, \vec{q}) M^*(t,\vec{x}, \vec{p}, -\vec{k}) \, ] \, 
									J(t,\vec{x},\vec{p}) ,
						 \label{eq:wke_gamma_nl} \\
			S_{\rm nl}[J,J](t, \vec{x},\vec{k}) 
						&	= 	\int \frac{\mathrm{d}^3 \vec{p} \, \mathrm{d}^3 \vec{q} }{(2 \pi)^3} \, 
									\delta^3(\vec{k} -\vec{p} -\vec{q}) \,
									\frac{\Theta(t,\vec{x},\vec{k},\vec{p},\vec{q}) }{	\mc{N} } \, |M(t,\vec{x}, \vec{p}, \vec{q})|^2\,
									J(t,\vec{x},\vec{p}) J(t,\vec{x},\vec{q}).
						 \label{eq:wke_source_nl} 	
		\end{align}
	\end{subequations}
\end{widetext}
Here $\Theta(t,\vec{x},\vec{k},\vec{p},\vec{q}) \doteq \pi \delta(\Delta \Omega)$, and
\begin{equation}
	\Delta \Omega(t,\vec{x},\vec{k},\vec{p},\vec{q})  \doteq \Omega(t,\vec{x},\vec{k})  - \Omega(t,\vec{x},\vec{p}) - \Omega(t,\vec{x},\vec{q}).
	\label{eq:wke_resonance}
\end{equation}
Also, 
\begin{equation}
	M(t,\vec{x},\vec{p},\vec{q})
		\doteq	M(t,\vec{x},p_0, \vec{p}, q_0, \vec{q}) |_{ p_0 = \Omega(t, \vec{x},\vec{p}), \, \, q_0 = \Omega(t, \vec{x},\vec{q}) },
\end{equation}
where
$M(x,p,q)$ is the scattering cross section 
\begin{equation}
	M(x,p,q)		\doteq	\alpha(x,p) \beta(x,q) + \alpha(x,q) \beta(x,p) ,
	\label{eq:wke_M}
\end{equation}
where $\alpha(z)$ and $\beta(z)$ are the Weyl symbols of the operators $\oper{\alpha}$ and $\oper{\beta}$, respectively. Note that $M(x,p,q)$ is symmetric with respect to its arguments $p$ and $q$. Finally, $\mc{N}(t,\vec{x},\vec{k}, \vec{p} ,\vec{q})$ is a normalization factor:
\begin{equation}
	\mc{N} \doteq	\pd_\omega D_{\rm H} (t,\vec{x},\vec{k}) \,
							\pd_\omega D_{\rm H} (t,\vec{x},\vec{p}) \,
							\pd_\omega D_{\rm H} (t,\vec{x},\vec{q}).
	\label{eq:wke_normalization}
\end{equation}

%%%%%%%%%%%%%%%%%%%%%%%%%%%%%%%%%%%%%%%%%%%%%%
\subsection{Discussion}

Equations \eq{eq:wke}--\eq{eq:wke_normalization} are the main result of our work. The WKE \eq{eq:wke} describes the wave-action density $J(t,\vec{x},\vec{k})$ in a weakly nonstationary and inhomogeneous medium with a quadratic nonlinearity. Concerning its linear dynamics, the left-hand side of \Eq{eq:wke} describes the Hamiltonian dynamics of $J(t,\vec{x},\vec{k})$ in the six-dimensional ray phase space. Physically, the left-hand side of \Eq{eq:wke} describes the wave refraction governed by the wave frequency $\Omega(t,\vec{x},\vec{k})$, which serves as a Hamiltonian for the system. On the right-hand side, $\gamma(t,\vec{x},\vec{k})$ defined in \Eq{eq:wke_gamma} represents the linear growth rate of the wave fluctuations and arises from the anti-Hermitian part of the modified dispersion operator $\oper{D}$ [\Eq{eq:basic_D_nl}]. The term $S_{\rm ext}(t,\vec{x},\vec{k})$ defined in \Eq{eq:wke_source} represents an external source term for the wave fluctuations.

\begin{figure}
	\includegraphics[scale=.5]{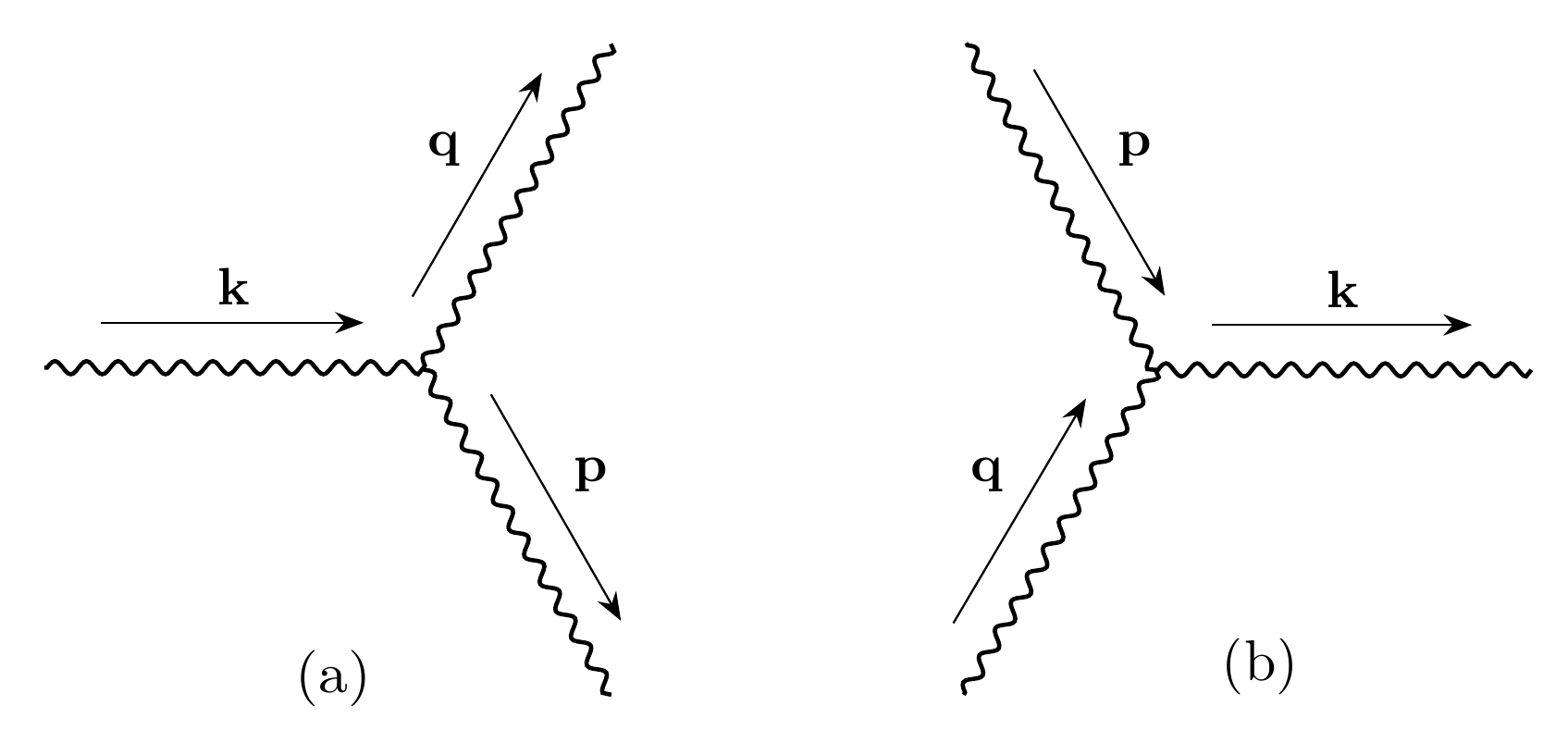}
	\caption{Feynman diagrams \cite{foot:diagrams} corresponding to the nonlinear wave-scattering processes. (a)~Coherent response $\gamma_{\rm nl}$: a wave $\vec{k}$ decays into two waves $\vec{p}$ and $\vec{q}$. (b)~Incoherent noise $S_{\rm nl}$: two waves $\vec{p}$ and $\vec{q}$ transfer their quanta into a wave $\vec{k}$.}
	\label{diagrams}
\end{figure}

The nonlinear term $C[J,J]$ in \Eq{eq:wke} plays the role of a \textit{wave scattering operator}. It is composed by two terms, $\gamma_{\rm nl} (t, \vec{x},\vec{k}) $ and $S_{\rm nl}(t, \vec{x},\vec{k})$, which arise from the nonlinear wave--wave interactions. The nonlinear source term $S_{\rm nl}$ in \eq{eq:wke_source_nl} is a bilinear functional on the wave action. It is always positive and represents contributions to the wave action $J(t,\vec{x},\vec{k})$ coming from waves with wavevectors $\vec{p}$ and $\vec{q}$ different from $\vec{k}$ (see \Fig{diagrams}). This term is also known as (the variance of) \textit{incoherent noise} \cite{Krommes:2002hva}. The nonlinear damping-rate term $\gamma_{\rm nl}$ in \Eq{eq:wke_gamma_nl} is linearly dependent of the wave action and represents a sink term where the wave action in the $\vec{k}$ wavevector is transferred to other modes with different wavevectors. The effects described by $\gamma_{\rm nl}$ are called the \textit{coherent response} \cite{Krommes:2002hva}. 

It is worth reminding that the WKE \eq{eq:wke} is coupled to \Eq{eq:stats_mean} for the mean field $\bar{\psi}(t,\vec{x})$. To lowest order in $\epsilon_{\rm go}$, \Eq{eq:stats_mean} is written in terms of the wave action as
\begin{align}
		(\oper{D}_{\rm lin} \bar{\psi}) (t,\vec{x}) 
					&	\simeq		\frac{1}{2} \,
								 f_{\rm nl}[\bar{\psi}, \bar{\psi} ](t,\vec{x}) 
								 +  \bar{ \mc{S} }(t,\vec{x})	\notag \\
					&		\quad 	 + \epsilon^2 \int \frac{ \mathrm{d}^3 \vec{k}}{(2 \pi)^3} \, \chi (t,\vec{x},\vec{k}) J(t,\vec{x},\vec{k})  ,
		\label{eq:wke_mean}
\end{align}
where we used \Eqs{eq:wke_ansatz} and \eq{eq:weyl_weyl_x_rep} and approximated the resulting Moyal products by ordinary products so that the integration kernel is
\begin{equation}
	\chi \doteq \left(	\frac{\alpha (t,\vec{x},\omega, \vec{k}) \beta (t,\vec{x},\omega, -\vec{k}) }{\pd D_{\rm H} / \pd \omega} \right)_{\omega=\Omega} (t,\vec{x},\vec{k}).
\end{equation}
One may recognize the last term in \Eq{eq:wke_mean} as a ponderomotive force. Then, one may also identify $\chi$ as the generalized susceptibility of the system with respect to fluctuations; cf. the so-called $K$-$\chi$ theorem \cite{Cary:1977ud,Kaufman:1987gh,Dodin:2017daa}

As a final note, since the mean field $\bar{\psi}$ is included in the definition of $\oper{D}$ in \Eq{eq:basic_D_nl}, whose Weyl symbol is used in deriving the WKE, then the mean field $\bar{\psi}$ will generally be present in the coefficients of the WKE \eq{eq:wke} and contribute to the dynamics of $J(t,\vec{x},\vec{k})$.

%%%%%%%%%%%%%%%%%%%%%%%%%%%%%%%%%%%%%%%%%%%%%%
%%%%%%%%%%%%%%%%%%%%%%%%%%%%%%%%%%%%%%%%%%%%%%
\section{Example}
\label{sec:example}

In the following, we shall consider a scalar wave $\psi(t,\vec{x})$ governed by the following equation:
\begin{equation}
	 \frac{\pd}{\pd t} \psi  + i H(t,\vec{x},-i \del) \psi =
			 \nu \del^2 \psi 
			+ \frac{\sigma}{2} \pd_x \psi^2,
	\label{eq:example_main}
\end{equation}
where the linear operator $H(t,\vec{x},-i \del) $ represents the linear frequency of the wave oscillations \cite{foot:example}, $\nu(t,\vec{x})$ is a small dissipation constant, and $\sigma(t,\vec{x})$ is a small function characterizing the quadratic nonlinearity similar to that found in the Burgers equation. This equation will serve as a useful example in which we can apply the general theory presented in \Sec{sec:WKE}.

As in \Sec{sec:mean_fluctuating}, we separate the wave field into its mean and fluctuating components. We obtain
\begin{subequations}
	\begin{gather}
		\pd_t \bar{\psi}
			+ i H(t,\vec{x},-i \del) \bar{\psi}
			 - \nu \del^2 \bar{\psi}= \frac{\sigma}{2} \pd_x ( \bar{\psi}^2 + \avg{ \fluctpsi^2 } ) , 
		\label{eq:example_mean}
	\end{gather}
	\begin{multline}
		\pd_t \fluctpsi
			+i H(t,\vec{x},-i \del) \fluctpsi
			 -  \nu \del^2 \fluctpsi 
			 -  \sigma ( \bar{\psi} \pd_x \fluctpsi +  \fluctpsi \pd_x \bar{\psi}  ) \\
			=  \frac{\sigma}{2} \pd_x ( \fluctpsi^2 -  \avg{ \fluctpsi^2 } )	.		
		\label{eq:example_fluct}
	\end{multline}
\end{subequations}
Following \Sec{sec:Dirac_abstract}, we write \Eq{eq:example_fluct} in the abstract vector representation. We first multiply \Eq{eq:example_fluct} by a factor $i$ so that
\begin{multline}
	i \pd_t \fluctpsi	-  H(t,\vec{x},-i \del) \fluctpsi
			 + i  \nu (-i\del)^2 \fluctpsi 
			 +  \sigma [ \bar{\psi} (-i \pd_x) \fluctpsi  \\ - i (\pd_x \bar{\psi})  \fluctpsi   ]
			=  \frac{\sigma}{2} (i\pd_x) ( \fluctpsi^2 -  \avg{ \fluctpsi^2 } )	.
	\label{eq:example_fluct_aux}
\end{multline}
In the spacetime representation, one has $\widehat{t}=t$, $\widehat{\vec{x}}=\vec{x}$, $\oper{\omega} = i \pd_t$, and $\oper{\vec{k}} = -i \del$. Hence, \Eq{eq:example_fluct_aux} can also be represented as in \Eq{eq:abstract_fluct}, where the modified dispersion operator is given by
\begin{equation}
	\oper{D} = 
			\widehat{\omega} 
			-\widehat{H}
			+ i \widehat{\nu} \widehat{\vec{k}}^2 
			+ \frac{\oper{\sigma}}{2} (  \widehat{\bar{\psi}} \, \widehat{k}_x - i \widehat{\bar{\psi}'} ) .
	\label{eq:example_oper_DNL}
\end{equation}
Here $\widehat{H} \doteq H(\widehat{t},\widehat{\vec{x}}, \widehat{\vec{k}})$ is the operator corresponding to the linear wave frequency. Likewise, $\widehat{\nu} \doteq \nu(\widehat{t},\widehat{\vec{x}})$ and $\widehat{\sigma} \doteq \sigma(\widehat{t},\widehat{\vec{x}})$. Also, $\widehat{\bar{\psi}} \doteq \bar{\psi}(\widehat{t},\widehat{\vec{x}})$ and $\widehat{\bar{\psi}'} \doteq (\pd_x \bar{\psi})(\widehat{t},\widehat{\vec{x}})$. (Here the tilde denotes a derivative on the $x$ coordinate.) 

Note that the nonlinear interaction term in \Eq{eq:example_fluct_aux} is $( \oper{\sigma} \psi ) ( -\widehat{k}_x \psi)$. Upon comparing to the primal equation \eq{eq:basic_original}, we find $\oper{\alpha} = \oper{\sigma}$ and $\oper{\beta} = -\widehat{k}_x$. After following the procedure in \Sec{app:construction}, we obtain the corresponding nonlinear wave--wave interaction operator appearing in \Eq{eq:abstract_fluct}:
\begin{equation}
	\oper{K}(x) =  \widehat{k}_x \ketlong{x} \bralong{x} \oper{\sigma} - \oper{\sigma} \ketlong{x} \bralong{x} \widehat{k}_x .
	\label{eq:example_M}
\end{equation}
This result can be verified by showing that
\begin{align}
	\frac{1}{2} \inner{ \fluctpsi \mid \oper{K}(x) \mid \fluctpsi }
			& 	=	\frac{1}{2} \inner{ \fluctpsi \mid \widehat{k}_x \mid x}  \inner{x \mid  \oper{\sigma} \mid \fluctpsi }	\notag \\
			& 		\quad -	\frac{1}{2}  \inner{ \fluctpsi \mid \oper{\sigma}  \mid x}  \inner{x \mid  \widehat{k}_x \mid \fluctpsi }	\notag \\
			& 	=	 \frac{1}{2} (-i \pd_x \fluctpsi)^*\,   \sigma \fluctpsi  
			 		-	\frac{1}{2} (  \sigma \fluctpsi )^*  (-i \pd_x \fluctpsi)	\notag \\
			& 	=	 i \frac{\sigma(t,\vec{x})}{2}    \pd_x \fluctpsi^2(t,\vec{x}).
	\label{eq:example_check}
\end{align}
Thus, the result in \Eq{eq:example_check} corresponds to the first term appearing in the right-hand side of \Eq{eq:example_fluct_aux}.

Following the procedure in \Sec{sec:WKE}, the next step is to calculate the Weyl transform \eq{eq:wke_weyl} of \Eq{eq:example_oper_DNL}. Using the Moyal product \eq{eq:weyl_Moyal} leads to the Weyl symbol of the nonlinear dispersion operator:
\begin{align}
	D 		&	=	\omega 
					- H(t,\vec{x},\vec{k})
					+ i \nu \star \vec{k}^2 
					+ \frac{\sigma}{2} \star \left( \bar{\psi} \star k_x - i \bar{\psi}' \right) \notag \\
			&	\simeq	\omega 
					- H(t,\vec{x},\vec{k})
					+ i \nu \vec{k}^2 
					+ \frac{\sigma}{2}  \left( \bar{\psi}  k_x - \frac{i}{2} \bar{\psi}' \right) ,
	\label{eq:example:Dnl_symbol}
\end{align}
where we used the fact that both $\nu$ and $\sigma$ are small so that we could replace the Moyal products by ordinary products. From \Eq{eq:example:Dnl_symbol}, the real and imaginary parts of the nonlinear dispersion symbol are given by
\begin{subequations}	\label{eq:example_real_imag}
	\begin{gather}
		D_{\rm H} (t,\vec{x},\omega, \vec{k}) 	= \omega  - H(t,\vec{x},\vec{k}) + \sigma \bar{\psi}  k_x /2 , \\
		D_{\rm A} (t,\vec{x},\vec{k})		= \nu \vec{k}^2 - \sigma \bar{\psi}' / 4.
	\end{gather}
\end{subequations}

From \Eqs{eq:stats_Wigner}, \eq{eq:example_mean}, and \eq{eq:weyl_weyl_x_rep}, we note that $\avg{\fluctpsi^2 } = (2 \pi)^{-4} \int \mathrm{d}^4 k \, W(x,k)$. Hence, after substituting \Eq{eq:wke_ansatz} and noting that $\pd_\omega D_H=1$, we obtain the governing equation for the mean field:
\begin{multline}
		 \pd_t \bar{\psi}
			+i H(t,\vec{x},-i \del) \bar{\psi}
			- \nu \del^2 \bar{\psi} \\
			=  \frac{\sigma}{2} \frac{ \pd \bar{\psi}^2}{\pd x} 
					+ \frac{\sigma}{2} \frac{\pd}{\pd x} \int \frac{\mathrm{d}^3 \vec{k}}{(2 \pi)^3} J(t,\vec{x},\vec{k}) , 	
	\label{eq:example_mean_result}
\end{multline}
As expected, the wave action $J(t,\vec{x},\vec{k})$ acts as a source term for the mean field.

Now, we shall derive the WKE for the fluctuations. Upon substituting \Eq{eq:example_real_imag} into \Eq{eq:wke}, we obtain
\begin{equation}
	\pd_t J + \{ J , \Omega \}
		=	2 \gamma J
			+C[J,J] ,
	\label{eq:example_WKE}
\end{equation}
where the wave frequency  and linear growth rate are
\begin{subequations}
	\begin{gather}
		\Omega(t,\vec{x},\vec{k} ) 		= H(t,\vec{x},\vec{k}) - \sigma   \bar{\psi}  k_x /2  , \\
		\gamma(t,\vec{x},\vec{k} ) 	  	= - \nu \vec{k}^2    + \sigma \bar{\psi}' /4.
	\end{gather}
\end{subequations}
For the terms concerning the nonlinear wave scattering $C= S_{\rm nl} [J,J]- 	2 \gamma_{\rm nl}[J]   J$, we note that the Weyl symbols entering $M(x,p,q)$ in \Eq{eq:wke_M} are $\alpha(z)=\sigma$ and $\beta(z)=-k_x$. Hence, it is clear that $M(t,\vec{x},\vec{p},\vec{q})=-\sigma ( p_x +q_x)$. Inserting this result into \Eqs{eq:wke_coeff_nl} leads to
\begin{subequations}
	\begin{align}
		\gamma_{\rm nl}(t,\vec{x},\vec{k})		& =  \sigma^2 k_x 
				\int \frac{\mathrm{d}^3 \vec{p}}{(2 \pi)^3}	   \,  \Theta_{\star} \, (k_x - p_x ) J(t,\vec{x},\vec{p}) ,
				\label{eq:example:gamma_nl} 	\\
		S_{\rm nl}(t,\vec{x},\vec{k})	&	= \sigma^2   k_x^2
				\int \frac{\mathrm{d}^3 \vec{p} \, }{(2 \pi)^3}		\,  \Theta_{\star} \, J(t,\vec{x},\vec{k}-\vec{p}) J(t,\vec{x},\vec{p}),
				\label{eq:example:F_nl} 
	\end{align}
\end{subequations}
where $\Theta_{\star} \doteq \pi\delta(\Delta \Omega_*$) and $\Delta \Omega_*$ is the frequency mismatch evaluated at $\vec{q}=\vec{k}-\vec{p}$; namely,
\begin{equation}
	\Delta \Omega_*
		\doteq	\Omega(t,\vec{x},\vec{k})  - \Omega(t,\vec{x},\vec{p}) - \Omega(t,\vec{x},\vec{k}-\vec{p}).
\end{equation}

%A few comments are warranted. It is to be noted that the frequency $\Omega(t,\vec{x},\vec{k} ) $ governing the wave Hamiltonian dynamics has a linear dependence on the mean field $\bar{\psi}$. Thus, the presence of the mean field can cause refraction on the incoherent waves. As expected, the term $- \nu \vec{k}^2$ represents linear damping of the wave action through conventional diffusion. The term $\sigma \bar{\psi}' /4$ induces linear growth (or dissipation). Regarding the wave scattering terms, there is no clear physical interpretation of $\gamma_{\rm nl}(t,\vec{x},\vec{k})$ and $S_{\rm nl}(t,\vec{x},\vec{k})$. This is due to the fact that the nonlinearity in \Eq{eq:example_main} does not lead to any obvious conserved invariant.

%%%%%%%%%%%%%%%%%%%%%%%%%%%%%%%%%%%%%%%%%%%%%%
%%%%%%%%%%%%%%%%%%%%%%%%%%%%%%%%%%%%%%%%%%%%%%

\section{Conclusions}
\label{sec:conclusions}

In this work, we presented a systematic derivation of a wave kinetic equation describing the dynamics of an incoherent wave field propagating in a medium with a weak quadratic nonlinearity. The medium is allowed to be weakly nonstationary and inhomogeneous. Primarily based on the Weyl phase-space representation, our derivation makes use of the well-known ordering assumptions of geometrical optics and of a statistical closure based on the quasinormal approximation. The resulting wave kinetic equation \eq{eq:wke} describes the wave dynamics in the ray phase space. It captures linear effects, such as refraction, linear damping, and external sources, as well as nonlinear wave scattering.

This work could be extended in several directions. From a practical point of view, the developed theory could be applied to study specific systems of weak wave turbulence interacting with mean fields in nonstationary and inhomogeneous media. In this regard, one possibility is to study the spontaneous formation of zonal flows in drift-wave turbulence present in magnetic fusion plasmas \cite{Diamond:2005br,Fujisawa:2009jc,EUROfusionConsortium:2016bk,Trines:2005in,Parker:2013hy,Parker:2016eu}. The present theory would include new effects due to wave scattering that are not captured in the commonly adopted quasilinear approximation. This work will be presented in a separate publication.

Also, there are at least two other possible avenues of future theoretical research. First, the theory presented in this work deals with a scalar field. It would be interesting to expand this in order to describe multicomponent waves. Second, the present theory made use of the quasinormal approximation in order to provide a statistical closure to the equations. However, there exists other closures that are presently being used to study strong turbulence. Examples include the direct-interaction approximation \cite{Kraichnan:1959cf} (also known as the DIA), the realizable Markovian closure \cite{Bowman:1992wta,Bowman:1993ia} (also known as the RMC), and the Martin--Sigga--Rose formalism \cite{Martin:1973zz} (also known as the MSR formalism). Hence, another possible avenue of research could be to marry the phase-space techniques presented here with the more advanced statistical closures previously mentioned. This could lead to new descriptions of weak (and possibly strong) wave turbulence in nonstationary and inhomogeneous media.

%Such a theory could be useful for describing more complex wave turbulence that exist in plasmas, such as the interaction of electromagnetic waves with Langmuir waves. 

%%%%%%%%%%%%%%%%%%%%%%%%%%%%%%%%%%%%%%%%%%%%%%
%%%%%%%%%%%%%%%%%%%%%%%%%%%%%%%%%%%%%%%%%%%%%%

\section*{Acknowledgments}

The authors would like to thank J. B. Parker, H. Zhu, and Y. Zhou for valuable discussions. This work was supported by the U.S. DOE through Contract DE-AC02-09CH11466, by the NNSA, and by Sandia National Laboratories. Sandia National Laboratories is a multimission laboratory managed and operated by National Technology and Engineering Solutions of Sandia, LLC., a wholly owned subsidiary of Honeywell International, Inc., for the U.S. DOE National Nuclear Security Administration under contract DE-NA-0003525.

%%%%%%%%%%%%%%%%%%%%%%%%%%%%%%%%%%%%%%%%%%%%%%
%%%%%%%%%%%%%%%%%%%%%%%%%%%%%%%%%%%%%%%%%%%%%%
\appendix

%%%%%%%%%%%%%%%%%%%%%%%%%%%%%%%%%%%%%%%%%%%%%%
%%%%%%%%%%%%%%%%%%%%%%%%%%%%%%%%%%%%%%%%%%%%%%

\section{Weyl calculus}
\label{app:weyl}

This appendix summarizes the conventions used for the Weyl symbol calculus. For more information, see the excellent reviews in \Refs{Ruiz:2017wc,Tracy:2014to,Imre:1967fr,BakerJr:1958bo,McDonald:1988dp}.

\par Let $\oper{A}$ be a linear operator. The Weyl symbol $A(x,k)$ is defined as the Weyl transform $\msf{W}[ \oper{A} ]$ of the linear operator $\oper{A}$; namely,
\begin{equation}
	A(x,k) \doteq \msf{W} [ \, \oper{A}	\, ] =
				\int \mathrm{d}^4 s \, e^{i k \cdot s }
				\inner{ x+ \tfrac{1}{2}s \mid \oper{A}	\mid x	-	\tfrac{1}{2}s	}  ,
	\label{eq:weyl_weyl_symbol}
\end{equation}
where $\ketlong{x}$ are the eigenstates of the position operator $\widehat{x}^\mu$ such that $\inner{x \mid \widehat{x}^\mu \mid x'} = x^\mu \delta^4(x-x')$. Also, $s\doteq(\tau,\vec{s})$, $k \cdot s = \omega \tau - \vec{k}\cdot \vec{s}$, and $\mathrm{d}^4 s \doteq \mathrm{d}\tau \, \mathrm{d}^3 \vec{s}$. The integrals span $\mathbb{R}^4$. This description of the operators is known as the \textit{phase-space representation} since the Weyl symbols are functions of the eight-dimensional phase space $z \doteq (x,k) = (t,\vec{x},\omega, \vec{k})$. Conversely, the inverse Weyl transform is given by
\begin{equation}
	\oper{A} = 
					\int	\frac{\mathrm{d}^4 x \,  \mathrm{d}^4 k \,  	\mathrm{d}^4 s }{(2 \pi)^4} \,
					e^{i k \cdot s } 
					A(x,k) \ketlong{x-\tfrac{1}{2}s} \bralong{x+\tfrac{1}{2}s}.
	\label{eq:weyl_weyl_inverse}
\end{equation}
The projection of the operator $\oper{A}$ on the position eigenstates, $\mcu{A}(x,x') = \inner{x \mid \oper{A}\mid x'}$, is
\begin{equation}
	\mcu{A}(x,x') =  
					\int	\frac{\mathrm{d}^4 k}{(2 \pi)^4} \,
					e^{-i k \cdot (x-x')  }
					A		\left(	\frac{x+x'}{2},	k	\right)  .
	\label{eq:weyl_weyl_x_rep}
\end{equation}

\par In the following, we shall outline a number of useful properties of the Weyl transform.
\begin{itemize}[leftmargin=*]
\item From \Eq{eq:weyl_weyl_x_rep}, we deduce that the trace $\mathrm{Tr} (\oper{A}) \doteq \int \mathrm{d}^4 x \, \inner{ x \mid  \oper{A} \mid  x }$ of an operator $\oper{A}$ is given by
\begin{equation}
	\mathrm{Tr} (\oper{A}) 
				=  \int \frac{\mathrm{d}^4 x \,  \mathrm{d}^4 k }{(2 \pi)^4}\,  
					A(x, k).
	\label{eq:weyl_trace}
\end{equation}

\item If $A(x,k)$ is the Weyl symbol of $\oper{A}$, then $A^*(x,k)$ is the Weyl symbol of $\oper{A}^\dag$. As a corollary, if the operator $\oper{A}$ is Hermitian $(\oper{A} = \oper{A}^\dag)$, then $A(z)$ is real.

\item For linear operators $\oper{A}$, $\oper{B}$ and $\oper{C}$ where $\oper{C} =\oper{A} \oper{B}$, the corresponding Weyl symbols satisfy
\begin{equation}
	C (x,k) = A (x,k) \star B (x,k).
	\label{eq:weyl_Moyal}
\end{equation}
Here ``$\star$" refers to the \textit{Moyal product} \citep{Moyal:1949gj}, which is given by
\begin{equation}
		A(x,k) \star B (x,k) 
		\doteq 
		A (x,k) e^{ i  \overleftrightarrow{\mc{L}}/ 2 }  B (x,k).
	\label{eq:weyl_Moyal2}
\end{equation}
Also, $\overleftrightarrow{\mc{L}}$ is the \textit{Janus operator}
\begin{equation}
	\overleftrightarrow{\mc{L}} =
			\frac{\overleftarrow{\pd} }{\pd \vec{x}}  \cdot \frac{ \overrightarrow{\pd} }{\pd \vec{k} }
	-		\frac{\overleftarrow{\pd} }{\pd \vec{k}}  \cdot \frac{ \overrightarrow{\pd} }{\pd \vec{x} }	
	+		\frac{\overleftarrow{\pd} }{\pd \omega}  \frac{ \overrightarrow{\pd} }{\pd t}
	-		\frac{\overleftarrow{\pd} }{\pd t}  \frac{ \overrightarrow{\pd} }{\pd \omega}.
	\label{eq:weyl_poisson_braket}
\end{equation}
The arrows indicate the direction in which the derivatives act. Note that $A \overleftrightarrow{\mc{L}}B = \{ A, B \}_8$ is the canonical Poisson bracket in the extended eight-dimensional phase space $(t,\vec{x},\omega, \vec{k})$.

\item The Moyal product is associative; \ie for arbitrary symbols $A$, $B$, and $C$, one has
\begin{equation}
	A \star B \star C 
			= (A \star B) \star C 
			= A \star (B \star C).
	\label{eq:weyl_associative}
\end{equation}

\item The anti-symmetrized Moyal product defines the so-called \textit{Moyal bracket}, namely,
\begin{equation}
	\moysin{ A ,	B } 
	 		\doteq -i	\left(  A \star B - B \star A 	\right)
	 					= 	2 A \sin \bigg( \frac{\overleftrightarrow{\mc{L}}}{2} \bigg) B .
	\label{eq:weyl_sine_bracket}
\end{equation}
Likewise, the symmetrized Moyal product is defined as
\begin{gather}
	 \moycos{ A , B} \doteq A \star B + B \star A 
	 								=	2 A \cos	\bigg(	\frac{\widehat{\mc{L}}}{2}		\bigg)  B.
 	\label{eq:weyl_cosine_bracket}
\end{gather}
Because of the latter equalities, these brackets are also referred as the \textit{sine} and \textit{cosine} brackets, respectively.

\item For fields that vanish rapidly enough at infinity,
\begin{gather}
 	\int \mathrm{d}^4 x \, \mathrm{d}^4 k \, 	A \star B 
   		= \int \mathrm{d}^4 x \, \mathrm{d}^4 k \, 		A B . 
 	\label{eq:weyl_int_everywhere}
\end{gather}

\item Now we give the Weyl transforms of various operators. The Weyl transform of the identity operator is
\begin{equation}
	\msf{W} [ \, \widehat{1}	\, ] =	1.
\end{equation}
The Weyl transforms of the time and position operators are given by
\begin{equation}
	\msf{W} [ \, \widehat{t} \, ] =	t, 		 \quad	\quad	\msf{W} [ \, \widehat{\vec{x}}  \, ] = \vec{x}.
\end{equation}
Likewise, the Weyl transforms of the frequency and wavevector operators are
\begin{equation}
	\msf{W} [ \, \widehat{\omega}  ] =	\omega, 		 \quad	\quad	\msf{W} [ \, \widehat{\vec{k}}  ] = \vec{k},
\end{equation}
where $\hat{\omega} = i\pd_t$ and $\hat{\vec{k}}=-i \del$ in the spacetime representation. For any two operators $f(\widehat{t}, \widehat{\vec{x}})$ and $g(\widehat{\omega},\widehat{\vec{k}})$,
\begin{equation}
	\msf{W} [\, f(\widehat{t}, \widehat{\vec{x}}) 	\, ]	=   f(t,\vec{x}),  \quad	\quad
	\msf{W} [\, g(\widehat{\omega},\widehat{\vec{k}}) \,]=	 g(\omega, \vec{k}).
\end{equation}
Upon using the Moyal product \eq{eq:weyl_Moyal2}, one can deduce
\begin{gather}
	\msf{W} [\,	\widehat{\vec{k}}	f(\widehat{\vec{x}}) 	\, ]	
								= \vec{k} f(\vec{x})	- \tfrac{i}{2} \del 	f(\vec{x}), \\
	\msf{W} [\, 	f(\widehat{\vec{x}}) \widehat{\vec{k}} \, ]	
								= \vec{k} f(\vec{x})	+ \tfrac{i}{2} \del 	f(\vec{x}).
\end{gather}

\end{itemize}

%%%%%%%%%%%%%%%%%%%%%%%%%%%%%%%%%%%%%%%%%%%%%%
%%%%%%%%%%%%%%%%%%%%%%%%%%%%%%%%%%%%%%%%%%%%%%
\section{Auxiliary calculations}
\label{app:auxiliary}

%%%%%%%%%%%%%%%%%%%%%%%%%%%%%%%%%%%%%%%%%%%%%%
\subsection{Constructing the operator $\boldsymbol{\oper{K} (x)}$}
\label{app:construction}

In the abstract Hilbert space, the nonlinear term appearing $f_{\rm nl}[\phi,\psi](x)$ in \Eq{eq:basic_f_nl} can be written as
\begin{align}
	f_{\rm nl}[\phi,\psi](x)
		=	& \, \inner{x \mid \oper{\alpha} \mid \phi} \inner{x \mid \oper{\beta} \mid \psi}	\notag \\
			& + \inner{x \mid \oper{\beta} \mid \phi} \inner{x \mid \oper{\alpha} \mid \psi} .
\end{align}
Now, let us introduce the transpose $\oper{A}_{\rm T}$ of an operator $\oper{A}$ such that $\inner{x \mid \oper{A}_{\rm T}\mid x'} \doteq \inner{ x' \mid \oper{A} \mid x}$;  in terms of components \cite{foot:transpose}, $\mcu{A}_{\rm T}(x,x') \doteq \mcu{A}(x',x)$. Examples of the transpose of commonly used operators are
\begin{gather*}
	\widehat{t}_{\rm T}=\widehat{t}	, 						\quad \quad 		\widehat{\vec{x}}_{\rm T}=\widehat{\vec{x}} , \quad \quad
	\widehat{\omega}_{\rm T}=-\widehat{\omega},	\quad \quad 	\widehat{\vec{k}}_{\rm T}=-\widehat{\vec{k}},
\end{gather*}
which can be immediately verified from the definitions in \Sec{sec:Dirac_abstract}. Hence, the transpose leaves the time and position operators intact but changes the sign of the frequency and wavevector operators. As in matrices, $(\oper{A} \oper{B})_{\rm T}= \oper{B}_{\rm T}\oper{A}_{\rm T}$; for example, $(\widehat{\vec{x}} \widehat{\vec{k}})_{\rm T}= \widehat{\vec{k}}_{\rm T}\widehat{\vec{x}}_{\rm T}= - \widehat{\vec{k}}  \widehat{\vec{x}} $.

If we consider $\phi(x)$ and $\psi(x)$ to be real, then \Eq{eq:basic_f_nl} can be written as
\begin{align}
	f_{\rm nl}[\phi,\psi](x)
		=	& \, \inner{\phi \mid \oper{\alpha}_{\rm T}\mid x} \inner{x \mid \oper{\beta} \mid \psi}	\notag \\
			& + \inner{\phi \mid \oper{\beta}_{\rm T}\mid x} \inner{x \mid \oper{\alpha} \mid \psi} .
\end{align}
From this result, we have $f_{\rm nl}[\phi,\psi](x) = \inner{\phi \mid \oper{K}(x) \mid \psi}$, where $\oper{K}(x)$ is given by \Eq{eq:abstract_K}.

%%%%%%%%%%%%%%%%%%%%%%%%%%%%%%%%%%%%%%%%%%%%%%
\subsection{Simplifying the Moyal products}
\label{app:simplification}

To derive the WKE \eq{eq:wke}, we need to approximate the Moyal products appearing in \Eq{eq:wke_wigner_moyal}. The most difficult terms to approximate are those involving derivatives of the Dirac delta functions. As an example, in this appendix we are interested in calculating the integral
\begin{equation}
	\mc{I} = \int \mathrm{d}\omega	\, A(z) \star W(z),
	\label{simp:eq:integral_orig}
\end{equation}
where $A(z)$ is an arbitrary function, $W(z)= 2 \pi \delta \boldsymbol{(} D_{\rm H} (z) \boldsymbol{)} J(t,\vec{x},\vec{k})$ is the GO ansatz \eq{eq:wke_ansatz}, and $z\doteq(t,\vec{x},\omega, \vec{k})$. The GO dispersion relation is $D_{\rm H}(z)=0$. As discussed in \Sec{sec:wke_derivation}, we shall assume a single dispersion branch such that $\omega = \Omega(t,\vec{x},\vec{k})$  satisfies this relation. Substituting \Eq{eq:wke_ansatz} into \Eq{simp:eq:integral_orig} leads to
\begin{align}
	\mc{I} 
		&	= 2 \pi \int \mathrm{d}\omega \,  A(z) \star [\delta(D_{\rm H}) J] \notag \\
		&	= 2 \pi \int \mathrm{d}\omega \, A(z) \exp\left( \frac{i}{2} \overleftrightarrow{\mc{L}} \right)  [\delta(D_{\rm H}) J] \notag \\
		& =2\pi \sum_{n=0}^\infty \frac{1}{n!} \left(\frac{i}{2}\right)^n \mc{I}_n, 
	\label{simp:eq:integral}
\end{align}
where $\overleftrightarrow{\mc{L}}$ is the Janus operator \eq{eq:weyl_poisson_braket} and
\begin{equation}
	\mc{I}_n \doteq \int d\omega\,A(z)	\left( \overleftrightarrow{\mc{L}} \right)^n [\delta(D_{\rm H}) J].
\end{equation}
%
%and $\overleftrightarrow{\mc{L}} \doteq \overleftarrow{\pd_\mu} \mc{J}^{\mu \nu} \overrightarrow{\pd_\nu}$, and $\mc{J}^{\mu \nu}$ is the canonical Poisson tensor in $z$ space.

Now, let us calculate each of the terms appearing in \Eq{simp:eq:integral}. The $n=0$ term is simply given by
\begin{align}
	\mc{I}_0 
		&	=	\int \mathrm{d}\omega \, A(z) \delta(D_{\rm H}) J	\notag \\
		&	=	A \boldsymbol{(} t,\vec{x},\Omega (t,\vec{x},\vec{k}) , \vec{k} \boldsymbol{)} \, \varrho^{-1}(t,\vec{x},\vec{k}) \, J(t,\vec{x},\vec{k}),
\end{align}
where the following function is independent of $\omega$:
\begin{equation}
	\varrho(t,\vec{x},\vec{k}) \doteq \left( \frac{\pd D_{\rm H}}{\pd \omega} \right)_{\omega = \Omega(t,\vec{x},\vec{k})} (t,\vec{x},\vec{k}).
\end{equation}

Let us now calculate the $n=1$ term in \Eq{simp:eq:integral}. After writing the Janus operator as $\overleftrightarrow{\mc{L}} \doteq \overleftarrow{\pd_\mu} \mc{J}^{\mu \nu} \overrightarrow{\pd_\nu}$, where $\mc{J}^{\mu \nu}$ is the canonical Poisson tensor in $z$ space, we obtain
\begin{widetext}
\begin{align}
	\mc{I}_1 
		&	\doteq 	\int \mathrm{d}\omega \, (\pd_\mu A) \mc{J}^{\mu \nu} \pd_\nu[  \delta(D_{\rm H}) J	] \notag \\
		&	=	\int \mathrm{d}\omega \, \frac{\pd A}{\pd z^\mu}  \mc{J}^{\mu \nu} 
				\frac{\pd}{\pd z^\nu}\left[ \delta(\omega-\Omega)  \varrho^{-1} J 	\right] \notag \\
		&	=	\int \mathrm{d}\omega \, \frac{\pd A}{\pd z^\mu} \mc{J}^{\mu \nu}  
				\left[ \frac{\pd \delta(\omega-\Omega)}{\pd z^\nu}  \varrho^{-1} J	
				+ \delta(\omega-\Omega)  \frac{\pd }{\pd z^\nu} \left( \varrho^{-1} J  \right)  \right] \notag \\
		&	=	\int \mathrm{d}\omega \, \frac{\pd A}{\pd z^\mu} \mc{J}^{\mu \nu}  
				\left[ \delta'(\omega-\Omega) \frac{\pd ( \omega-\Omega)}{\pd z^\nu}  \varrho^{-1} J
				+ \delta(\omega-\Omega)  \frac{\pd }{\pd z^\nu} \left( \varrho^{-1} J \right)  \right] \notag \\
		&	=	\int \mathrm{d}\omega \,  
				\left[ - \frac{\pd }{\pd \omega} \left(  \frac{\pd A}{\pd z^\mu}   \mc{J}^{\mu \nu}  
				\frac{\pd ( \omega-\Omega)}{\pd z^\nu}  \varrho^{-1} J \right) 
				+  \frac{\pd A}{\pd z^\mu} \mc{J}^{\mu \nu} \frac{\pd }{\pd z^\nu} \left( \varrho^{-1} J \right)  \right] \delta(\omega-\Omega)  \notag \\
		&	=	\left[ -  \frac{\pd }{ \pd z^\mu }\left( \frac{\pd A}{\pd \omega}\right)    \mc{J}^{\mu \nu}  
				\frac{\pd ( \omega-\Omega)}{\pd z^\nu}  \varrho^{-1} J
				+  \frac{\pd A}{\pd z^\mu} \mc{J}^{\mu \nu} \frac{\pd }{\pd z^\nu} \left( \varrho^{-1} J  \right)  \right]_{\omega=\Omega}.
	\label{simp:eq:calculation}
\end{align}
Here we integrated by parts in $\omega$ and used the fact that $J$, $\Omega$ and $\varrho$ is independent of $\omega$. Upon explicitly writing the derivatives in terms of phase-space coordinates, we have 
\begin{equation}
	\mc{I}_1
			=	\varrho^{-1} J
				\left( \frac{\pd \Omega}{\pd t}  \frac{\pd^2 A}{\pd^2 \omega}  
							+ \frac{\pd^2 A}{\pd t \pd \omega } 
							- \left\{ \Omega, \frac{\pd A}{\pd \omega}\right\}_6 \right)_{\omega=\Omega}  
		 		+  \varrho^{-1}  \left( \frac{\pd A }{\pd \omega}  \frac{\pd J}{\pd t} + \{ A, J\}_6 \right)_{\omega=\Omega} 
				-   \varrho^{-2} J
							\left( 	\frac{\pd A}{\pd \omega} \frac{\pd \varrho}{\pd t}  + 
										\left\{ A,  \varrho \right\} \right)_{\omega=\Omega} ,
\end{equation}
where $\{ \cdot , \cdot \}_6 \doteq \overleftarrow{\pd_\vec{x}} \cdot \overrightarrow{\pd_\vec{k}} - \overleftarrow{\pd_\vec{k}} \cdot \overrightarrow{\pd_\vec{x}}$ is the six-dimensional Poisson bracket. We then rearrange the terms so that
\begin{align}
	\mc{I}_1
		&	=	\varrho^{-1} J
				\left( \frac{\pd \Omega}{\pd t}  \left( \frac{\pd^2 A}{\pd^2 \omega}   \right)_{\omega=\Omega} 
							+ \left( \frac{\pd^2 A}{\pd t \pd \omega }  \right)_{\omega=\Omega}  
							- \frac{\pd }{\pd t} \left[ \left( \frac{\pd A }{\pd \omega} \right)_{\omega=\Omega} \right]  \right)  \notag \\
		&	\quad 		+  \varrho^{-1}  \left( \frac{\pd A }{\pd \omega}  \frac{\pd J}{\pd t} + \{ A, J\}_6 \right)_{\omega=\Omega} 
							-  \varrho^{-1} J
							\left(	\left\{ \Omega, \frac{\pd A}{\pd \omega}\right\}_6 
									+ \varrho^{-1} 
										\left\{ A, \varrho    \right\}_6 \right)_{\omega=\Omega} .
\end{align}
\end{widetext}
The terms in the first line cancel. Hence,
\begin{align}
	\mc{I}_1
		&	=	\varrho^{-1}  \left( \frac{\pd A }{\pd \omega}  \frac{\pd J}{\pd t} + \{ A, J\}_6 \right)_{\omega=\Omega} \notag \\
		&		\quad 		-  \varrho^{-1} J
							\left(	\left\{ \Omega, \frac{\pd A}{\pd \omega}\right\}_6 
									+ \varrho^{-1} 
										\left\{ A, \varrho    \right\}_6 \right)_{\omega=\Omega} .
	\label{eq:integral_I}
\end{align}

As can be seen, the derivatives appearing in \Eq{eq:integral_I} are all acting on smooth functions. Hence, $\mc{I}_1 = \mc{O}(\epsilon_{\rm go})$ in the GO limit, where $\epsilon_{\rm go}$ is the GO parameter \eq{eq:wke_epsilon}. By reiterating the same calculation given in \Eq{simp:eq:calculation}, one can show that the higher $n$ terms in \Eq{simp:eq:integral} scale as $\mc{I}_n = \mc{O}(\epsilon_{\rm go}^n)$. With these results and using the fact that the Moyal brackets [\Eqs{eq:weyl_sine_bracket} and \eq{eq:weyl_cosine_bracket}] are defined in terms of Moyal products, one can simplify \Eq{eq:wke_wigner_moyal} in order to obtain the WKE \eq{eq:wke}.

In particular, let us calculate the special case where $A = D_H$. Upon substituting the GO relations \cite{Whitham:2011kb}
\begin{subequations}
	\begin{gather}
		\frac{\pd \Omega }{\pd \vec{x} }		(t,\vec{x},\vec{k})
						=	-\left(	\frac{\pd D_{\rm H} / \pd \vec{x} }{ \pd D_{\rm H} / \pd \omega	} \right)_{\omega=\Omega}  (t,\vec{x},\vec{k}), \\
		\frac{\pd \Omega }{\pd \vec{k} }		(t,\vec{x},\vec{k})
						=  - \left(	\frac{\pd D_{\rm H} / \pd \vec{k} }{ \pd D_{\rm H} / \pd \omega	} \right)_{\omega=\Omega} (t,\vec{x},\vec{k}) ,
	\end{gather}
\end{subequations}
and using \Eq{eq:integral_I}, we first note that
\begin{align}
	\int & \mathrm{d}\omega \, D_{\rm H} \overleftrightarrow{\mc{L}} [ \delta(D_{\rm H}) J	] 	\notag \\
		&	=	 \pd_t J +  \{ J , \Omega \}_6  
				-  \varrho^{-1}
							\left[	\left( \left\{ \Omega, \pd_\omega D_{\rm H} \right\}_6  \right)_{\omega=\Omega}
									-	\left\{ \Omega, \varrho   \right\}_6 \right].
	\label{simp:eq:integral_I_example}
\end{align}
The last term in \Eq{simp:eq:integral_I_example} gives zero:
\begin{align}
	\big(  & \left\{ \Omega,  \pd_\omega D_{\rm H}  \right\}_6   \big)_{\omega=\Omega} 
									-	\left\{ \Omega, \varrho   \right\}_6 	\notag \\
		&	=	\pd_\vec{x} \Omega \cdot \left( \pd_\vec{k} \pd_\omega D_{\rm H} \right)_{\omega=\Omega}
				-	\pd_\vec{k} \Omega \cdot \left( \pd_\vec{x} \pd_\omega D_{\rm H} \right)_{\omega=\Omega}	\notag \\
		&	\quad		
				- \pd_\vec{x} \Omega \cdot \pd_\vec{k} \varrho
				+ \pd_\vec{k} \Omega \cdot \pd_\vec{x} \varrho
				\notag \\
		&	=	-\pd_\vec{x} \Omega \cdot \pd_\vec{k} \Omega \left( \pd^2_\omega D_{\rm H} \right)_{\omega=\Omega} 
				+\pd_\vec{k} \Omega \cdot \pd_\vec{x} \Omega  \left( \pd^2_\omega D_{\rm H} \right)_{\omega=\Omega}  \notag \\
		&	=0.
\end{align}
Hence, we obtain
\begin{equation}
	\int  \mathrm{d}\omega \, D_{\rm H} \overleftrightarrow{\mc{L}} [ \delta(D_{\rm H}) J	] 
		=  \pd_t J+  \{ J , \Omega \}_6  .
	\label{simp:eq:advection}
\end{equation}
which is the advection term appearing in the WKE \eq{eq:wke}.

%%%%%%%%%%%%%%%%%%%%%%%%%%%%%%%%%%%%%%%%%%%%%%
\subsection{Calculation of $\boldsymbol{\eta(z)}$ and $\boldsymbol{F(z)}$}
\label{app:nonlinear}

We begin by calculating the Weyl symbol $F(z)$ corresponding to $\oper{F}$ in \Eq{eq:closure_F}. Upon substituting \Eq{eq:abstract_K}, we first note that the trace appearing in \Eq{eq:closure_F} can be written as

\begin{widetext}
\begin{align}
	\mathrm{Tr}[ \,  \oper{K}(x) \oper{W}  \oper{K}^\dag (y) \oper{W} \, ] 
		=		&	\quad	\inner{x \mid \oper{\beta} \,  \oper{W}	 \, \oper{\alpha}^\dag		\mid y } 
							\inner{y \mid \oper{\beta}_{\rm T}^\dag \, \oper{W} \, \oper{\alpha}_{\rm T}\mid x }	 
				 		+	\inner{x \mid \oper{\beta} \, \oper{W} \, \oper{\beta}^\dag 	\mid y }
							 \inner{y \mid \oper{\alpha}_{\rm T}^\dag \,	\oper{W} \,	\oper{\alpha}	_{\rm T}\mid x }	 \notag \\
				& 		+	\inner{x \mid \oper{\alpha} \, \oper{W} \, \oper{\alpha}^\dag   \mid y }  
							\inner{y\mid \oper{\beta}_{\rm T}^\dag \,  \oper{W}\,  \oper{\beta}_{\rm T} \mid x }	
				 		+	\inner{x \mid \oper{\alpha} \, \oper{W} \, \oper{\beta}^\dag  \mid y } 
							\inner{y \mid \oper{\alpha}_{\rm T}^\dag \, \oper{W} \, \oper{\beta}_{\rm T} \mid x }	 .
	\label{eq:nonlinear_trace}
\end{align}
After substituting \Eq{eq:nonlinear_trace} into \Eq{eq:closure_F} and applying the Weyl transform \eq{eq:wke_weyl}, we obtain integrals of the form
\begin{equation*}
	\int \mathrm{d}^4 s \, e^{i k \cdot s} \,	\inner{x+\tfrac{1}{2}s \mid \oper{A} \mid x-\tfrac{1}{2}s} 
																\inner{x-\tfrac{1}{2}s \mid \oper{B} \mid x+\tfrac{1}{2}s},
\end{equation*}
where $\oper{A}$ and $\oper{B}$ are given by the terms appearing in \Eq{eq:nonlinear_trace}. To evaluate these integrals, we can use the identity
\begin{equation}
	\int \mathrm{d}^4 s \, e^{i k \cdot s} \,	\inner{x+\tfrac{1}{2}s \mid \oper{A} \mid x-\tfrac{1}{2}s} 
																\inner{x-\tfrac{1}{2}s \mid \oper{B} \mid x+\tfrac{1}{2}s}
		=	\int \frac{\mathrm{d}^4 p  \, \mathrm{d}^4 q	  }{(2\pi)^4} \, \delta^4(k-p-q) A(x,p) B(x,-q).
	\label{eq:nonlinear_identity}
\end{equation}
This property is analogous to the convolution theorem frequently used in Fourier transforms. Hence, we have
\begin{align}
	F(x,k)	 =		\frac{1}{2}		\int \frac{\mathrm{d}^4 p  \, \mathrm{d}^4 q	  }{(2\pi)^4} \, \delta^4(k-p-q)
							\big\{ & \, \, \, \,
								(\beta \star W \star \alpha^*)(x,p) ([\beta_{\rm T}]^* \star W \star [\alpha_{\rm T}] )(x,-q) 	\notag \\
				& 			+ (\beta \star W \star \beta^*)(x,p) ( [\alpha_{\rm T}]^* \star W \star [\alpha_{\rm T}] ) (x,-q) 	\notag \\
				& 			+ (\alpha \star W \star \alpha^*) (x,p)  ([\beta_{\rm T}]^* \star W \star \beta_{\rm T})(x,-q)		\notag \\
				& 			+ ( \alpha \star W \star \beta^* )(x,p) ( [\alpha_{\rm T}]^* \star W \star \beta_{\rm T})(x,-q)  \, \, \, \,  \big\},
	\label{eq:nonlinear_F_aux}
\end{align}							 
where we used the Moyal product \eq{eq:weyl_Moyal}. Here $\alpha(x,k)$ and $\beta(x,k)$ are the Weyl symbols corresponding to $\oper{\alpha}$ and $\oper{\beta}$, respectively. Also, $[\alpha_{\rm T}](x,k)$ denotes the Weyl symbol of $\oper{\alpha}_{\rm T}$. Equation \eq{eq:nonlinear_F_aux} can be further simplified as follows. Since we consider the wave nonlinearities to be weak, the Moyal products can be replaced by ordinary products. Moreover, for an arbitrary operator $\oper{A}$, one has $[A_{\rm T}](x,k) = \msf{W}[\oper{A}_{\rm T}] = A(x,-k)$ by using \Eq{eq:weyl_weyl_symbol}. Hence, we obtain
\begin{equation}
	F(x,k)		=		\frac{1}{2}	
							\int \frac{\mathrm{d}^4 p  \, \mathrm{d}^4 q	  }{(2\pi)^4} \, \delta^4(k-p-q)
							|M(x,p,q)|^2 W(x,p) W(x,q) 
							+ \mc{O}(\epsilon_{\rm go}),
	\label{eq:nonlinear_F}
\end{equation}
where $W(x,k)$ is given by \Eq{eq:wke_ansatz} and we used the reality property so $W(x,q) = W(x,-q)$. We also introduced
\begin{equation}
	M(x,p)		\doteq	\alpha(x,p) \beta(x,q) + \alpha(x,q) \beta(x,p) .
	\label{eq:nonlinear_M}
\end{equation}
Finally, when substituting \Eq{eq:nonlinear_F} into $S_{\rm nl}(t,\vec{x},\vec{k})$ in \Eq{eq:wke_source_nl_aux}, we obtain the result reported in \Eq{eq:wke_source_nl}.

Now, let us compute the Weyl symbol $\eta(z)$ of $\oper{\eta}$ in \Eq{eq:closure_eta}. Substituting \Eq{eq:abstract_K} into \Eq{eq:closure_eta} leads to
\begin{align}
	\oper{\eta}
		=		- \int \mathrm{d}^4 x \, & \mathrm{d}^4 y 
					\ketlong{x} \bralong{y}
					(\oper{D}^{-1})^\dag \, 
				 \big[  
				 		\quad \oper{\alpha}_{\rm T}\ketlong{x} \bralong{x} \oper{\beta} \, \oper{W} \, \oper{\alpha}^\dag \ketlong{y} \bralong{y} \oper{\beta}_{\rm T}^\dag	
								\quad 	+  \quad
						\oper{\alpha}_{\rm T}\ketlong{x} \bralong{x} \oper{\beta} \, \oper{W} \, \oper{\beta}^\dag \ketlong{y} \bralong{y} \oper{\alpha}_{\rm T}^\dag
						\notag \\
			&		+ 	\oper{\beta}_{\rm T}\ketlong{x} \bralong{x} \oper{\alpha} \, \oper{W} \, \oper{\alpha}^\dag \ketlong{y} \bralong{y} \oper{\beta}_{\rm T}^\dag	
								\quad 	+ \quad      
						\oper{\beta}_{\rm T}\ketlong{x} \bralong{x} \oper{\alpha} \, \oper{W} \, \oper{\beta}^\dag \ketlong{y} \bralong{y} \oper{\alpha}_{\rm T}^\dag 
					\quad	\big] .
	\label{eq:nonlinear_eta_aux}
\end{align}
To calculate the Weyl transform of the above, we shall use the following result:
\begin{align}
	\msf{W} \left[ \int \mathrm{d}^4 u \,  \mathrm{d}^4 v  \,
								\ketlong{u} 
								\inner{ v \mid \oper{A} \mid u} \inner{u \mid \oper{B} \mid v} \bralong{v} \oper{C} \right]	
		&	= \left( \int \mathrm{d}^4 s \, e^{i k \cdot s} \,	\inner{x-\tfrac{1}{2}s \mid \oper{A} \mid x+\tfrac{1}{2}s} 
																\inner{x+\tfrac{1}{2}s \mid \oper{B} \mid x-\tfrac{1}{2}s} 			\right) \star C(x,k)
		\notag \\
		&	=	\left(  \int \frac{\mathrm{d}^4 p  \, \mathrm{d}^4 q	  }{(2\pi)^4} \, \delta^4(k-p-q) A(x,-q) B(x,p) \right) \star C(x,k) ,
	\label{eq:nonlinear_identity_II}
\end{align}
where in the first line, we decomposed the Weyl transform using the Moyal product, and in the second line, we used the identity in \Eq{eq:nonlinear_identity}. By using this identity, we calculate the Weyl transform of \Eq{eq:nonlinear_eta_aux}. Similarly as in \Eq{eq:nonlinear_F}, we then approximate the resulting Moyal products with ordinary products. To leading order, we obtain
\begin{equation}
	\eta(x,k)
		=	-		\int \frac{ \mathrm{d}^4 p \, \mathrm{d}^4 q}{(2 \pi)^4} \, 
					\delta^4(k -p - q) 
					[D^{-1}]^*(x,-q) M(x,p,q) M^*(x,p,-k)  
					W(x,p) 
					+ \mc{O}(\epsilon_{\rm go}),
	\label{eq:nonlinear_eta}
\end{equation}
where $W(x,k)$ is given by \Eq{eq:wke_ansatz}. Finally, substituting \Eq{eq:nonlinear_eta} into $\eta_{\rm nl}(t,\vec{x},\vec{k})$ in \Eq{eq:wke_gamma_nl_aux} leads to the result reported in \Eq{eq:wke_gamma_nl}.

\end{widetext}

%%%%%%%%%%%%%%%%%%%%%%%%%%%%%%%%%%%%%%%%%%%%%%


\begin{thebibliography}{69}
\bibitem[{foo({\natexlab{a}})}]{foot:nonlinear_interaction}
\bibinfo{note}{More generally, the nonlinearity can involve a convolution in
  time or space. This general case will not be considered here.}
\bibitem[{foo({\natexlab{b}})}]{foot:real_psi}
\bibinfo{note}{We shall not discuss in detail the necessary conditions so that
  the solution of \Eq{eq:basic_original} is real. We shall simply take as given
  that $\psi(x)$ is real.}
\bibitem[{foo({\natexlab{c}})}]{foot:wavevector}
\bibinfo{note}{In the spacetime representation, we have the following: $\inner{
  x \mid \widehat{\vec{k}} \mid \psi } = \int \mathrm{d}^4 x' \, \inner{ x \mid
  \widehat{\vec{k}} \mid x' } \inner{ x' \mid \psi } = - i \del_\vec{x} \int
  \mathrm{d}^4 x' \, \delta^4(x -x') \psi(x') = -i \del \psi(t,\vec{x})$.}
\bibitem[{foo({\natexlab{e}})}]{foot:reality}
\bibinfo{note}{Note that in order to satisfy the condition $W(x,k)=W(x,-k)$,
  the Hermitian part of the nonlinear dispersion operator must satisfy $\delta
  \boldsymbol{(} D_{\rm H}(x,k) \boldsymbol{)} = \delta \boldsymbol{(} D_{\rm
  H}(x,-k) \boldsymbol{)}$, which implies that the dispersion operator must
  satisfy $D_{\rm H}(x,k)=D_{\rm H}(x,-k)$ or $D_{\rm H}(x,k)=-D_{\rm
  H}(x,-k)$.}
\bibitem[{foo({\natexlab{f}})}]{foot:inverse}
\bibinfo{note}{As a reminder, $[D_{\rm nl}^{-1}]$ and $D_{\rm nl}^{-1}$ are not
  the equivalent. $[D_{\rm nl}^{-1}]$ is the Weyl symbol of $\oper{D}_{\rm
  nl}^{-1}$ while $D_{\rm nl}^{-1}$ is simply the inverse of the Weyl symbol
  $D_{\rm nl}$ corresponding to $\oper{D}_{\rm nl}$. In general, $[D_{\rm
  nl}^{-1}] \neq D_{\rm nl}^{-1}$; however, note that the Weyl representation
  of $\oper{D}_{\rm nl} \oper{D}_{\rm nl}^{-1} = \widehat{1}$ is $D_{\rm nl}
  \star [ D_{\rm nl}^{-1} ]= 1$. By replacing the Moyal product with an
  ordinary product, we obtain $[D_{\rm nl}^{-1}] \simeq D_{\rm nl}^{-1}$.}
\bibitem[{foo({\natexlab{g}})}]{foot:Plemelj}
\bibinfo{note}{The Sokhotski--Plemelj theorem states that if $f(x)$ is function
  which is continuous on the real line, then $\lim_{\varepsilon \to 0^+}
  \integral \mathrm{d}x \, \frac{f(x)}{x\pm i \epsilon} = \mp i \pi f(0) +
  \mc{P} \integral \mathrm{d}x \, \frac{f(x)}{x} $, where $\mc{P}$ denotes the
  Cauchy principal value.}
\bibitem[{foo({\natexlab{i}})}]{foot:example}
\bibinfo{note}{For $\psi$ to be real, the coefficients appearing in
  \Eq{eq:example_main} must be real. Hence, one necessarily needs $i
  H(t,\vec{x},-i \del) = \mc{H}(t,\vec{x}, \del)$. We chose to write this
  equation in this form in order to explicitly show the Hermitan momentum
  operator $\widehat{\vec{k}}=-i \del$.}
\bibitem[{foo({\natexlab{j}})}]{foot:transpose}
\bibinfo{note}{Note that the transpose is similarly defined as the transpose of
  matrices in linear algebra.}
\end{thebibliography}
\end{document}